\documentclass[12pt,hyperfootnotes=false]{article}
\usepackage{mathptmx}
\usepackage[latin9]{inputenc}
\usepackage{geometry}
\usepackage{float}
\usepackage{textcomp}
\usepackage{amsmath}
\usepackage{amsthm}
\usepackage{amssymb}
\usepackage{graphicx}
\usepackage{setspace}
\usepackage[authoryear]{natbib}
\usepackage[unicode=true]
 {hyperref}
\usepackage{breakurl}
\usepackage{array}
\usepackage{tablefootnote}
\usepackage{longtable}
\usepackage{multirow}
\makeatletter

\DeclareTextSymbolDefault{\textquotedbl}{T1}
\providecommand{\tabularnewline}{\\}

\usepackage{amsthm}
\usepackage{bm}
\usepackage{setspace}
\usepackage{sectsty}

\usepackage{datetime}
\usepackage{pdflscape}
\usepackage[english]{babel}
\usepackage{times}
\usepackage[small]{caption}
\usepackage[bottom,hang,flushmargin]{footmisc}

\sectionfont{\noindent\normalfont\large\bf}
\subsectionfont{\noindent\normalfont\normalsize\bf}
\subsubsectionfont{\noindent\normalfont\it}

\makeatother

\begin{document}

\title{\textbf{Trends in the Diffusion of Misinformation }\\
\textbf{on Social Media}}

\author{\noindent Hunt Allcott,\textbf{ }\emph{New York University, Microsoft
Research, and NBER}\textbf{}\thanks{E-mail: hunt.allcott@nyu.edu, gentzkow@stanford.edu, chuanyu@stanford.edu.
We thank the Stanford Institute for Economic Policy Research (SIEPR),
the Stanford Cyber Initiative, the Toulouse Network for Information
Technology, the Knight Foundation, and the Alfred P. Sloan Foundation
for generous financial support. We thank David Lazer, Brendan Nyhan,
David Rand, David Rothschild, Jesse Shapiro, and Nils Wernerfelt for
helpful comments and suggestions. We also thank our dedicated research
assistants for their contributions to this project.}\textbf{}\\
Matthew Gentzkow, \textit{Stanford University and NBER}\textbf{}\\
Chuan Yu,\textbf{ }\emph{Stanford University}\textbf{}\\
}

\date{\monthname\ \number\year}
\maketitle
\begin{abstract}
\noindent We measure trends in the diffusion of misinformation on
Facebook and Twitter between January 2015 and July 2018. We focus
on stories from 570 sites that have been identified as producers of
false stories. Interactions with these sites on both Facebook and
Twitter rose steadily through the end of 2016. Interactions then fell
sharply on Facebook while they continued to rise on Twitter, with
the ratio of Facebook engagements to Twitter shares falling by approximately
60 percent. We see no similar pattern for other news, business, or
culture sites, where interactions have been relatively stable over
time and have followed similar trends on the two platforms both before
and after the election.

\bigskip{}
\end{abstract}
\pagebreak{}

\begin{spacing}{1.2}

\section{Introduction}

Misinformation on social media has caused widespread alarm in recent
years. A substantial number of U.S. adults were exposed to false news
stories prior to the 2016 election, and post-election surveys suggest
that many people who read such stories believed them to be true (Silverman
and Singer-Vine 2016; Allcott and Gentzkow 2017; Guess et al. 2018).
Many argue that false news stories played a major role in the 2016
election (for example, Olson 2016; Parkinson 2016; Read 2016; Gunther
et al. 2018), and in the ongoing political divisions and crises that
have followed it (for example, Spohr 2017; Azzimonti and Fernandes
2018; Tharoor 2018). Numerous efforts have been made to respond to
the threat of false news stories, including educational and other
initiatives by civil society organizations, hearings and legal action
by regulators, and a range of algorithmic, design, and policy changes
made by Facebook and other social media companies.

Evidence on whether these efforts have been effective---or how the
scale of the misinformation problem is evolving more broadly---remains
limited. A recent study argues that false stories remain a problem
on Facebook even after changes to its news feed algorithm in early
2018 (Newswhip 2018). The study reports that the 26th and 38th most
engaging stories on Facebook in the two months after the changes were
from fake news websites. Many articles that have been rated as false
by major fact-checking organizations have not been flagged in Facebook\textquoteright s
system, and two major fake news sites have seen little or no decline
in Facebook engagements since early 2016 (Funke 2018). Facebook's
now-discontinued strategy of flagging inaccurate stories as \textquotedblleft Disputed\textquotedblright{}
can modestly lower the perceived accuracy of flagged headlines (Blair
et al. 2017), though some research suggests that the presence of warnings
can cause untagged false stories to be seen as more accurate (Pennycook
and Rand 2017). Media commentators have argued that efforts to fight
misinformation through fact-checking are \textquotedblleft not working\textquotedblright{}
(Levin 2017) and that misinformation overall is \textquotedblleft becoming
unstoppable\textquotedblright{} (Ghosh and Scott 2018).

In this paper, we present new evidence on the volume of misinformation
circulated on social media from January 2015 to July 2018. We assemble
a list of 570 sites identified as sources of false stories in a set
of five previous studies and online lists. We refer to these collectively
as \emph{fake news sites}. We measure the volume of Facebook engagements
and Twitter shares for all stories on these sites by month. As points
of comparison, we also measure the same outcomes for stories on (i)
a set of major news sites; (ii) a set of small news sites not identified
as producing misinformation; and (iii) a set of sites covering business
and culture topics.

The results show that interactions with the fake news sites in our
database rose steadily on both Facebook and Twitter from early 2015
to the months just after the 2016 election. Interactions then declined
by more than half on Facebook, while they continued to rise on Twitter.
The ratio of Facebook engagements to Twitter shares was roughly steady
at around 40:1 from the beginning of our period to late 2016, then
fell to roughly 15:1 by the end of our sample period. In contrast,
interactions with major news sites, small news sites, and business
and culture sites have all remained relatively stable over time, and
have followed similar trends on Facebook and Twitter both before and
after the 2016 election. While this evidence is far from definitive,
we see it as consistent with the view that the overall magnitude of
the misinformation problem may have declined, at least temporarily,
and that efforts by Facebook following the 2016 election to limit
the diffusion of misinformation may have had a meaningful impact

The results also show that the absolute level of interaction with
misinformation remains high, and that Facebook continues to play a
particularly important role in its diffusion. In the period around
the election, fake news sites received almost as many Facebook engagements
as the 38 major news sites in our sample. Even after the sharp drop
following the election, Facebook engagements of fake news sites still
average roughly 70 million per month.

Our evidence is subject to many important caveats and must be interpreted
with caution. This is particularly true for the raw trends in interactions.
While we have attempted to make our database of false stories as comprehensive
as possible, it is likely far from complete, and many factors could
generate selection biases that vary over time. The raw decline in
Facebook engagements may partly reflect the under-sampling of sites
that could have entered or gained popularity later in our sample period,
as well as efforts by producers of misinformation to evade detection
on Facebook by changing their domain names. It may also reflect changes
over time in demand for highly partisan political content that would
have existed absent efforts to fight misinformation, and could reverse
in the future, for example in the run-up to future elections.

We see the comparison of Facebook engagements to Twitter shares as
potentially more informative. If the design of these platforms and
the behavior of their users were stable over time, we might expect
sample selection biases or demand changes to have similar proportional
effects, and thus leave the ratio of Facebook engagements to Twitter
shares roughly unchanged. For example, we might expect producers changing
domain names to evade detection to produce similar declines in our
measured interactions on both platforms. The fact that Facebook engagements
and Twitter shares follow similar trends prior to late 2016 and for
the non-fake-news sites in our data, but diverge sharply for fake
news sites following the election, suggests that some factor has slowed
the relative diffusion of misinformation on Facebook. The suite of
policy and algorithmic changes made by Facebook following the election
seems like a plausible candidate.

However, even the relative comparison of the platforms is only suggestive.
Both Facebook and Twitter have made changes to their platforms, and
so at best this measure captures the relative effect of the former
compared to the latter. Engagements on Facebook affect sharing on
Twitter and vice versa. The selection of stories into our database
could for various reasons differentially favor the kinds of stories
likely to be shared on one platform or the other, and this selection
could vary over time. Demand changes need not have the same proportional
effect on the two platforms. Some of these factors would tend to attenuate
changes in the Facebook-Twitter ratio, leading our results to be conservative,
but others could produce a spurious decrease over time.

In the appendix, we show that our qualitative results survive a set
of robustness checks intended to partially address potential sample
selection biases. These checks include: (i) focusing on sites identified
as fake in multiple lists; (ii) excluding sites from each of our five
lists in turn, (iii) looking at sites that were active in different
periods; (iv) excluding potential outliers and looking at sites of
different sizes; and (v) looking at sites with different likelihoods
to publish misinformation.

\section{Background}

Both Facebook and Twitter have taken steps to reduce the circulation
of misinformation on their platforms. In the appendix, we list twelve
such announcements by Facebook and five by Twitter since the 2016
election. Broadly, the platforms have taken three types of actions
to limit misinformation. First, they have limited its supply, by blocking
ads from pages that repeatedly share false stories and removing accounts
that violate community standards. Second, they have introduced features
such as \textquotedblleft disputed\textquotedblright{} flags or \textquotedblleft related
articles\textquotedblright{} that provide corrective information related
to a false story. Third, they have changed their algorithms to de-prioritize
false stories in favor of news from trustworthy publications and posts
from friends and family.

Legislators are also taking action. For example, Connecticut, New
Mexico, Rhode Island, and Washington passed laws in 2017 encouraging
media literacy and digital citizenship (Zubrzycki 2017). Executives
from Facebook, Google, and Twitter have been asked to testify before
various congressional committees about their efforts to combat misinformation
(Shaban et al. 2017; Popken 2018). Although there has been no major
national legislation, this testimony may have raised public awareness.

Finally, civil society organizations also play an important role.
For example, the News Literacy Project provides non-partisan educational
materials to help teachers educate students to evaluate the credibility
of information; demand for its materials has grown substantially in
the past few years (Strauss 2018). In 2017, the newly established
News Integrity Initiative (NII) made ten grants totaling \$1.8 million
to help build trust between newsrooms and the public, make newsrooms
more diverse and inclusive, and make public conversations less polarized
(Owen 2017).

\section{Data}

We compile a list of sites producing false news stories by combining
five previous lists: (i) a research project by Grinberg et al. (2018,
490 sites); (ii) PolitiFact's article titled \textquotedblleft PolitiFact's
guide to fake news websites and what they peddle\textquotedblright{}
(Gillin 2017, 325 sites); (iii) three articles by BuzzFeed on fake
news (Silverman 2016; Silverman et al. 2017a; Silverman et al. 2017b;
223 sites); (iv) a research project by Guess et al. (2018, 92 sites);
and (v) FactCheck's article titled \textquotedblleft Websites that
post fake and satirical stories\textquotedblright{} (Schaedel 2017,
61 sites). Politifact and FactCheck are independent journalistic fact-checking
websites, while BuzzFeed similarly applies journalistic standards
to evaluating whether articles are true or false. The two lists from
research projects originally derive from subsets of the other three,
plus Snopes.com, another independent fact-checking site, and lists
assembled by blogger Brayton (2016) and media studies scholar Zimdars
(2016). The union of these five lists is our set of fake news sites.

PolitiFact and FactCheck work directly with Facebook to evaluate the
veracity of stories flagged by Facebook users as potentially false.
Thus, these lists comprise fake news sites that Facebook is likely
to be aware are fake. As a result, our results may be weighted toward
diffusion of \emph{misinformation that Facebook is aware of}, and
may not fully capture trends in \emph{misinformation that Facebook
is not aware of}. It is difficult to assess how large this latter
group might be. Our list almost certainly includes the most important
providers of false stories, as Facebook users can flag any and all
questionable articles for review. On the other hand, the list might
exclude a large tail of small sites producing false stories.

Combining these five lists yields a total of 673 unique sites. We
have data for 570 of them. We report in the appendix the names and
original lists of 50 largest sites in terms of total Facebook engagements
plus Twitter shares during the sample period. In our robustness checks,
we consider alternative rules for selecting the set of sites.

Our sets of comparison sites are defined based on category-level web
traffic rankings from Alexa.\footnote{\copyright 2018, Alexa Internet (www.alexa.com)}
Alexa measures web traffic using its global traffic panel, a sample
of millions of Internet users who have installed browser extensions
allowing their browsing data to be recorded, plus data from websites
that use Alexa to measure their traffic. It then ranks sites based
on a combined measure of unique visitors and pageviews. We define
major news sites to be the top 100 sites in Alexa's News category.
We define small news sites to be the sites ranked 401-500 in the News
category. We define business and culture sites to be the top 50 sites
in each of the Arts, Business, Health, Recreation, and Sports categories.
For each of these groups, we omit from our sample government websites,
databases, sites that do not mainly produce news or similar content,
international sites whose audiences are primarily outside the U.S.,
and sites that are included in our list of fake news sites. Our final
sample includes 38 major news sites, 78 small news sites, and 54 business
and culture sites.

We gather monthly Facebook engagements and Twitter shares of all articles
published on these sites from January 2015 to July 2018 from BuzzSumo
(www.buzzsumo.com). BuzzSumo is a commercial content database that
tracks the volume of user interactions with internet content on Facebook,
Twitter, and other social media platforms, using data available from
the platforms' application programming interfaces (APIs). We use BuzzSumo's
data on total Facebook engagements and total Twitter shares by originating
website and month. Facebook engagements are defined as the sum of
shares, comments, and reactions such as \textquotedbl likes.\textquotedbl{}
We have data for 570 out of 673 fake news sites in our list and all
sites in the comparison groups. We sum the monthly Facebook engagements
and Twitter shares of articles from all sites in each category and
then average by quarter.

\section{Results}

\begin{figure}[H]
\caption{Engagement on Facebook and Twitter\label{fig:trends}}

\medskip{}

\begin{centering}
\begin{tabular}{c}
\emph{Panel A: Facebook Engagements}\tabularnewline
\includegraphics[scale=0.85]{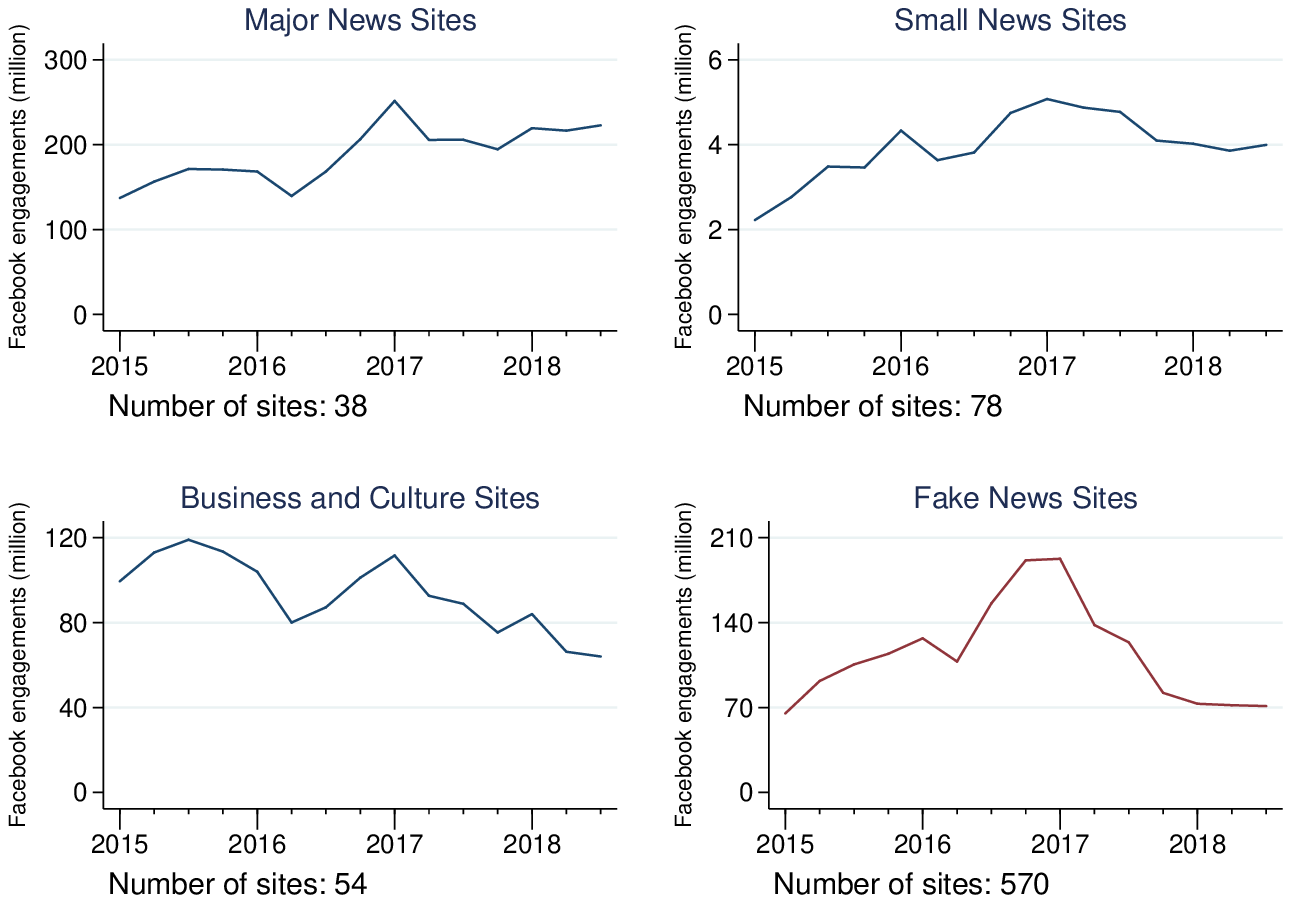}\tabularnewline
\emph{Panel B: Twitter Shares}\tabularnewline
\includegraphics[scale=0.85]{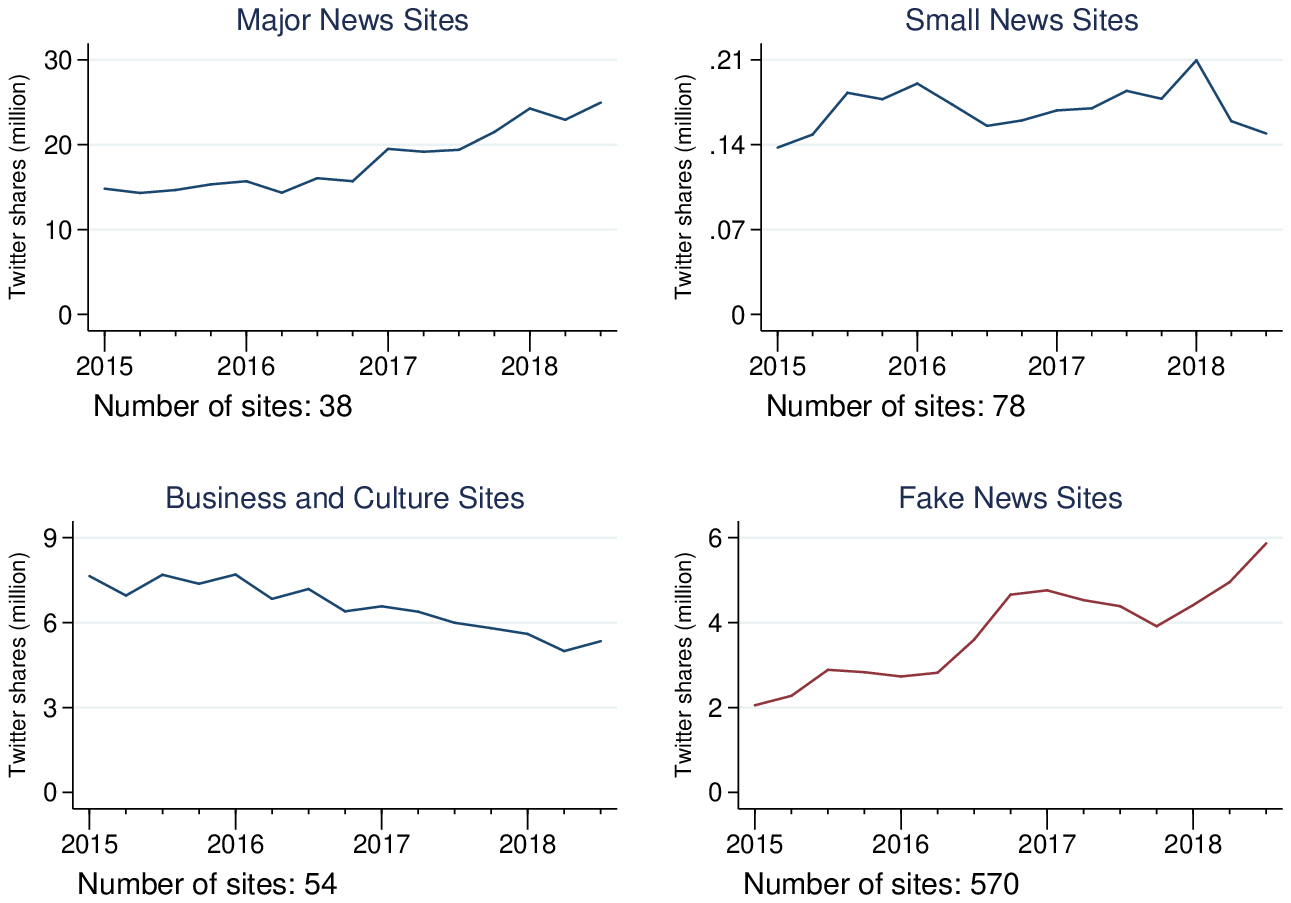}\tabularnewline
\end{tabular}
\par\end{centering}
\medskip{}

\emph{\footnotesize{}Notes: }{\footnotesize{}This figure shows monthly
Facebook engagements and Twitter shares of all articles published
on sites in different categories averaged by quarter. Data comes from
BuzzSumo. Major News Sites include 38 sites selected from the top
100 sites in Alexa's News category. Small News Sites include 78 sites
selected from the sites ranking 401-500 in the News category. Business
and Culture Sites include 54 sites selected from the top 50 sites
in each of the Arts, Business, Health, Recreation, and Sports categories.
Fake News Sites include 570 sites assembled from five lists. The complete
lists can be found in the appendix.}{\footnotesize\par}
\end{figure}

Figure \ref{fig:trends} shows trends in the number of Facebook engagements
and Twitter shares of stories from each category of site. Interactions
for major news sites, small news sites, and business and culture sites
have remained relatively stable during the past two years, and follow
similar trends on Facebook and Twitter. Both platforms show a modest
upward trend for major news and small news sites, and a modest downward
trend for business and culture sites. In contrast, interactions with
fake news have changed more dramatically over time, and these changes
are very different on the two platforms. Fake news interactions increased
steadily on both platforms from the beginning of 2015 up to the 2016
election. Following the election, however, Facebook engagements fell
sharply (declining by more than 50 percent), while shares on Twitter
continued to increase.

\begin{figure}[H]
\caption{Relative Engagement on Facebook\label{fig:ratio}}

\medskip{}

\begin{centering}
\begin{tabular}{c}
\includegraphics{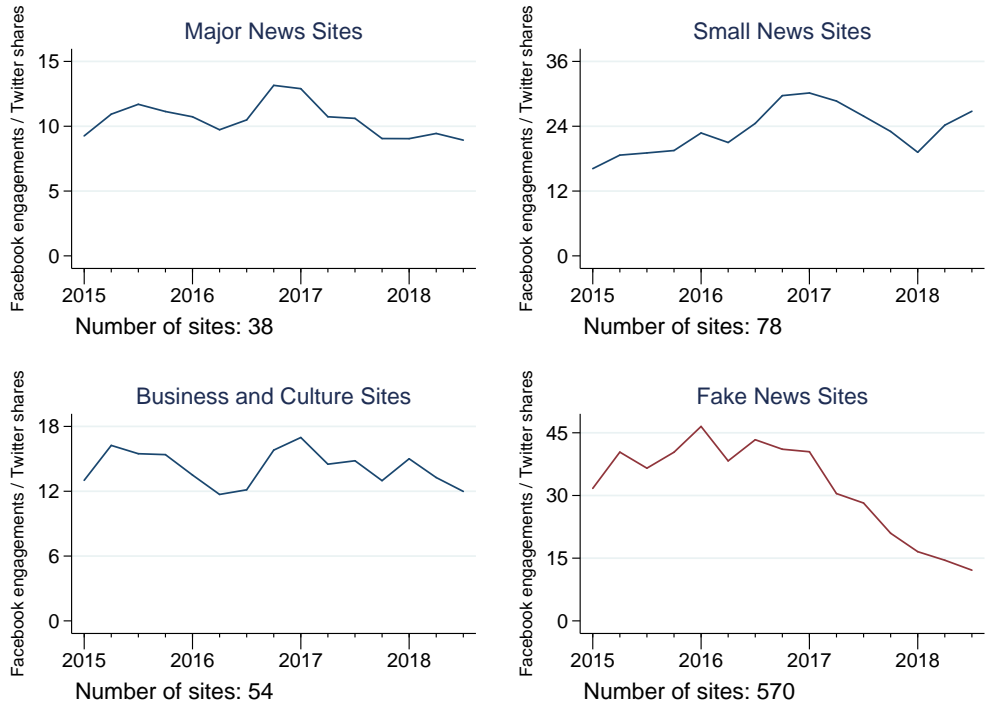}\tabularnewline
\end{tabular}
\par\end{centering}
\medskip{}

\emph{\footnotesize{}Notes: }{\footnotesize{}This figure shows the
ratio of monthly Facebook engagements over Twitter shares of all articles
published on sites in different categories averaged by quarter. Data
comes from BuzzSumo. Major News Sites include 38 sites selected from
the top 100 sites in Alexa's News category. Small News Sites include
78 sites selected from the sites ranking 401-500 in the News category.
Business and Culture Sites include 54 sites selected from the top
50 sites in each of the Arts, Business, Health, Recreation, and Sports
categories. Fake News Sites include 570 sites assembled from five
lists. The complete lists can be found in the appendix.}{\footnotesize\par}
\end{figure}

Figure \ref{fig:ratio} shows our main result: trends in the ratio
of Facebook engagements to Twitter shares. The ratios have been relatively
stable for major news, small news, and business and culture sites.
For fake news sites, however, the ratio has declined sharply, from
around 45:1 during the election to around 15:1 two years later.

While these results suggest that the circulation of fake news on Facebook
has declined in both absolute and relative terms, it is important
to emphasize that the absolute quantity of fake news interactions
on both platforms remains large, and that Facebook in particular has
played an outsized role in its diffusion. Figure \ref{fig:trends}
shows that Facebook engagements fell from a peak of roughly 200 million
per month at the end of 2016 to roughly 70 million per month at the
end of our sample period. As a point of comparison, the 38 major news
sites in the top left panel---including the New York Times, Wall
Street Journal, CNN, Fox News, etc.---typically garner about 200-250
million Facebook engagements per month. On Twitter, fake news shares
have been in the 4-6 million per month range since the end of 2016,
compared to roughly 20 million per month for the major news sites.

We report a number of robustness checks in the appendix designed to
address concerns about selection into our sample of sites. First,
we restrict to sites that are identified as fake news sites by at
least two or three of our original five lists, which leaves 116 and
19 sites, respectively. Second, given that people might disagree with
any one particular study's list of fake news sites, we run five additional
analyses, each excluding fake news sites identified exclusively by
one of our five lists. Third, we focus on sites that started active
operations after November 2016, sites that were still in active operation
as of July 2018, and sites that were in active operation from August
2015 to July 2018, which leaves 226, 215, and 82 sites respectively.
(Active operation is defined to be a global traffic rank reported
by Alexa of at least one million.) Fourth, we exclude the five largest
sites in terms of total interactions to ensure the trend is not driven
solely by outliers. We also look at sites in the first decile and
sites in the bottom nine deciles separately to see if the trend holds
for both large sites and small sites. Fifth, Grinberg et al. (2018)
provide three lists of sites classified by different likelihoods to
publish misinformation. We look at each of these lists separately.
Our main qualitative conclusions remain consistent across these checks,
though the exact size and shape of the trends vary. Finally, we present
an alternative comparison group: a small set of politically focused
sites such as Politico and The Hill. These sites do see a decline
in engagements on Facebook relative to Twitter, but it mainly occurred
in late-2015.\pagebreak{}

\newpage{}

\end{spacing}

\pagebreak{}

\section*{References}

\leftskip=2em
\parindent=-2em
\onehalfspacing

\noindent

Allcott H, Gentzkow M (2017) Social media and fake news in the 2016
election. \emph{Journal of Economic Perspectives} 31(2): 211--236.

Azzimonti M, Fernandes M (2018) Social media networks, fake news,
and polarization. \emph{NBER Working Paper No. 24462}.

Blair S, et al. (2017) Real solutions for fake news? Measuring the
effectiveness of general warnings and fact-check tags in reducing
belief in false stories on social media. \emph{Working Paper.} Available
at \href{https://www.dartmouth.edu/~nyhan/fake-news-solutions.pdf}{https://www.dartmouth.edu/$\sim$nyhan/fake-news-solutions.pdf}.
Accessed September 6, 2018.

Brayton E (2016) Please stop sharing links to these sites. \emph{Patheos}.
Available at \href{http://www.patheos.com/blogs/dispatches/2016/09/18/please-stop-sharing-links-to-these-sites/}{http://www.patheos.}\\
\href{http://www.patheos.com/blogs/dispatches/2016/09/18/please-stop-sharing-links-to-these-sites/}{com/blogs/dispatches/2016/09/18/please-stop-sharing-links-to-these-sites/}.
Accessed September 5, 2018.

Funke D (2018) Fact-checkers have debunked this fake news site 80
times. It\textquoteright s still publishing on Facebook. \emph{Poynter.org}.
Available at \href{https://www.poynter.org/news/fact-checkers-have-debunked-fake-news-site-80-times-its-still-publishing-facebook}{https://www.poynter.org/news/fact-checkers-have-de-}\\
\href{https://www.poynter.org/news/fact-checkers-have-debunked-fake-news-site-80-times-its-still-publishing-facebook}{bunked-fake-news-site-80-times-its-still-publishing-facebook}.
Accessed September 4, 2018.

Ghosh, D, Scott B (2018) Disinformation is becoming unstoppable. \emph{Time}.
Available at \href{http://time.com/5112847/facebook-fake-news-unstoppable/}{http://time.}\\
\href{http://time.com/5112847/facebook-fake-news-unstoppable/}{com/5112847/facebook-fake-news-unstoppable/}.
Accessed September 2, 2018.

Gillin J (2017) Politifact\textquoteright s guide to fake news websites
and what they peddle. \emph{PolitiFact}. Available at \href{http://www.politifact.com/punditfact/article/2017/apr/20/politifacts-guide-fake-news-websites-and-what-they/}{http://www.politifact.com/punditfact/article/2017/apr/20/politifacts-guide-fake-news-}\\
\href{http://www.politifact.com/punditfact/article/2017/apr/20/politifacts-guide-fake-news-websites-and-what-they/}{websites-and-what-they/}.
Accessed September 3, 2018.

Grinberg N, Joseph K, Friedland L, Swire-Thompson B, Lazer D (2018)
Fake news on Twitter during the 2016 U.S. presidential election. \emph{Working
Paper}. Available from the authors.

Guess A, Nyhan B, Reifler J (2018) Selective exposure to misinformation:
evidence from the consumption of fake news during the 2016 US presidential
campaign. \emph{Working Paper. European Research Council}. Available
at \href{https://www.dartmouth.edu/~nyhan/fake-news-2016.pdf}{https://www.dartmouth.edu/$\sim$nyhan/fake-news-2016.pdf}.
Accessed September 3, 2018.

Gunther R, Beck PA, Nisbet EC (2018) Fake news may have contributed
to Trump\textquoteright s 2016 victory. Available at \href{https://assets.documentcloud.org/documents/4429952/Fake-News-May-Have-Contributed-to-Trump-s-2016.pdf}{https://assets.documentcloud.org/documents/4429952/Fake-News-May-Have-}\\
\href{https://assets.documentcloud.org/documents/4429952/Fake-News-May-Have-Contributed-to-Trump-s-2016.pdf}{Contributed-to-Trump-s-2016.pdf}.
Accessed September 3, 2018.

Levin S (2017) Facebook promised to tackle fake news. But the evidence
shows it\textquoteright s not working. \emph{The Guardian}. Available
at \href{https://www.theguardian.com/technology/2017/may/16/facebook-fake-news-tools-not-working}{https://www.theguardian.com/technology/2017/may/16/facebook}\\
\href{https://www.theguardian.com/technology/2017/may/16/}{-fake-news-tools-not-working}.
Accessed September 3, 2018.

NewsWhip (2018) Navigating the Facebook algorithm change: 2018 report.
Available at \href{http://go.newswhip.com/rs/647-QQK-704/images/FacebookAlgorithmMarch18.pdf}{http://go.}\\
\href{http://go.newswhip.com/rs/647-QQK-704/images/FacebookAlgorithmMarch18.pdf}{newswhip.com/rs/647-QQK-704/images/FacebookAlgorithmMarch18.pdf}.
Accessed September 3, 2018.

Olson P (2016) How Facebook helped Donald Trump become president.
\emph{Forbes}. Available at \href{https://www.forbes.com/sites/parmyolson/2016/11/09/how-facebook-helped-donald-trump-become-president/}{https://www.forbes.com/sites/parmyolson/2016/11/09/how-facebook-helped-donald-trump-}\\
\href{https://www.forbes.com/sites/parmyolson/2016/11/09/how-facebook-helped-donald-trump-become-president/}{become-president/}.
Accessed September 3, 2018.

Owen LH (2017) The News Integrity Initiative gives \$1.8 million to
10 projects focused on increasing trust in news. \emph{NiemanLab}.
Available at \href{http://www.niemanlab.org/2017/10/the-news-integrity-initiative-gives-1-8-million-to-10-projects-focused-on-increasing-trust-in-news/}{http://www.niemanlab.org/2017/10/the-news-}\\
\href{http://www.niemanlab.org/2017/10/the-news-integrity-initiative-gives-1-8-million-to-10-projects-focused-on-increasing-trust-in-news/}{integrity-initiative-gives-1-8-million-to-10-projects-focused-on-increasing-trust-in-news/}.
Accessed September 3, 2018.

Parkinson HJ (2016) Click and elect: how fake news helped Donald Trump
win a real election. \emph{The Guardian}. Available at \href{https://www.theguardian.com/commentisfree/2016/nov/14/fake-news-donald-trump-election-alt-right-social-media-tech-companies}{https://www.theguardian.com/commentisfree/2016/nov/14/fake-}\\
\href{https://www.theguardian.com/commentisfree/2016/nov/14/fake-news-donald-trump-election-alt-right-social-media-tech-companies}{news-donald-trump-election-alt-right-social-media-tech-companies}.
Accessed September 2, 2018.

Pennycook G, Rand DG (2017) The implied truth effect: attaching warnings
to a subset of fake news stories increases perceived accuracy of stories
without warnings. \emph{Working Paper}. Available at \href{http://dx.doi.org/10.2139/ssrn.3035384}{http://dx.doi.org/10.2139/ssrn.3035384}.
Accessed September 3, 2018.

Popken B (2018) Apologies and promises: Facebook and Twitter tell
senators they will do more to combat misinformation. \emph{NBC News}.
Available at \href{https://www.nbcnews.com/tech/tech-news/apologies-promises-facebook-twitter-tell-senators-they-will-do-more-n906731}{https://www.nbcnews.com/tech/tech-news/}\\
\href{https://www.nbcnews.com/tech/tech-news/apologies-promises-facebook-twitter-tell-senators-they-will-do-more-n906731news/apologies-promises-facebook-twitter-tell-senators-they-will-do-more-n906731}{apologies-promises-facebook-twitter-tell-senators-they-will-do-more-n906731}.
Accessed \\
September 8, 2018.

Read M (2016) Donald Trump won because of Facebook. \emph{New York
Magazine}. Available at \href{http://nymag.com/selectall/2016/11/donald-trump-won-because-of-facebook.html}{http://nymag.com/selectall/2016/11/donald-trump-won-because-of-facebook.html}.
Accessed September 3, 2018.

Schaedel S (2017) Websites that post fake and satirical stories. \emph{FactCheck}.
Available at \href{https://www.factcheck.org/2017/07/websites-post-fake-satirical-stories}{https://www.}\\
\href{https://www.factcheck.org/2017/07/websites-post-fake-satirical-stories}{factcheck.org/2017/07/websites-post-fake-satirical-stories}.
Accessed September 3, 2018.

Shaban H, Timberg C, Dwoskin E (2017) Facebook, Google and Twitter
testified on Capitol Hill. Here\textquoteright s what they said. \emph{The
Washington Post}. Available at \href{https://www.washingtonpost.com/news/the-switch/wp/2017/10/31/facebook-google-and-twitter-are-set-to-testify-on-capitol-hill-heres-what-to-expect}{https://www.washingtonpost.com/}\\
\href{https://www.washingtonpost.com/news/the-switch/wp/2017/10/31/facebook-googleand-twitter-are-set-to-testify-on-capitol-hill-heres-what-to-expect}{news/the-switch/wp/2017/10/31/facebook-google-and-twitter-are-set-to-testify-on-capitol-hill}\\
\href{https://www.washingtonpost.com/news/the-switch/wp/2017/10/31/facebook-google-and-twitter-are-set-to-testify-on-capitol-hill-heres-what-to-expect}{-heres-what-to-expect}.
Accessed September 7, 2018

Silverman C (2016) Here are 50 of the biggest fake news hits on Facebook
from 2016. \emph{BuzzFeed News}. Available at \href{https://www.buzzfeednews.com/article/craigsilverman/top-fake-news-of-2016}{https://www.buzzfeednews.com/article/craigsilverman/top-fake-news-of-}\\
\href{https://www.buzzfeednews.com/article/craigsilverman/top-fake-news-of-2016}{2016}.
Accessed September 3, 2018.

Silverman C, Lytvynenko J, Pham S (2017a) These are 50 of the biggest
fake news hits on Facebook in 2017. \emph{BuzzFeed News}. Available
at \href{https://www.buzzfeednews.com/article/craigsilverman/these-are-50-of-the-biggest-fake-news-hits-on-facebook-in}{https://www.buzzfeednews.com/article/craigsilver}\\
\href{https://www.buzzfeednews.com/article/craigsilverman/these-are-50-of-the-biggest-fake-news-hits-on-facebook-in}{man/these-are-50-of-the-biggest-fake-news-hits-on-facebook-in}
Accessed September 3, 2018.

Silverman C, Singer-Vine J (2016) Most Americans who see fake news
believe it, new survey says. \emph{BuzzFeed News}. Available at \href{https://www.buzzfeed.com/craigsilverman/fake-news-survey}{https://www.buzzfeed.com/craigsilverman/fake-news-survey}.
Accessed September 3, 2018.

Silverman C, Singer-Vine J, Vo LT (2017b) In spite of the crackdown,
fake news publishers are still earning money from major ad networks.
\emph{BuzzFeed News}. Available at \href{https://www.buzzfeednews.com/article/craigsilverman/fake-news-real-ads}{https://www.buzzfeed}\\
\href{https://www.buzzfeednews.com/article/craigsilverman/fake-news-real-ads}{news.com/article/craigsilverman/fake-news-real-ads}.
Accessed September 3, 2018.

Spohr D (2017) Fake news and ideological polarization: filter bubbles
and selective exposure on social media. \emph{Business Information
Review} 34(3): 150--160.

Strauss V (2018) The News Literacy Project takes on \textquoteleft fake\textquoteright{}
news --- and business is better than ever. \emph{The Washington Post.}
Available at \href{https://www.washingtonpost.com/news/answer-sheet/wp/2018/03/27/not-sure-whats-real-or-fake-anymore-the-news-literacy-project-teaches-kids-how-to-tell-the-difference-and-its-growing-faster-than-ever/}{https://www.washingtonpost.com/news/answer-}\\
\href{https://www.washingtonpost.com/news/answer-sheet/wp/2018/03/27/not-sure-whats-real-or-fake-anymore-the-news-literacy-project-teaches-kids-how-to-tell-the-difference-and-its-growing-faster-than-ever/}{sheet/wp/2018/03/27/not-sure-whats-real-or-fake-anymore-the-news-literacy-project-teaches}\\
\href{https://www.washingtonpost.com/news/answer-sheet/wp/2018/03/27/not-sure-whats-real-or-fake-anymore-the-news-literacy-project-teaches-kids-how-to-tell-the-difference-and-its-growing-faster-than-ever/}{-kids-how-to-tell-the-difference-and-its-growing-faster-than-ever/}.
Accessed September 3,\\
2018.

Tharoor I (2018) \textquoteleft Fake news\textquoteright{} and the
Trumpian threat to democracy. \emph{The Washington Post}. Available
at \href{https://www.washingtonpost.com/news/worldviews/wp/2018/02/07/fake-news-and-the-trumpian-threat-to-democracy/}{https://www.washingtonpost.com/news/worldviews/wp/2018/02/07/fake-news-and-}\\
\href{https://www.washingtonpost.com/news/worldviews/wp/2018/02/07/fake-news-and-the-trumpian-threat-to-democracy/}{the-trumpian-threat-to-democracy/}.
Accessed September 3, 2018.

Zimdars M (2016) False, misleading, clickbait-y, and satirical \textquotedblleft news\textquotedblright{}
sources. Available at \href{http://d279m997dpfwgl.cloudfront.net/wp/2016/11/Resource-False-Misleading-Clickbaity-and-Satirical-\%E2\%80\%9CNews\%E2\%80\%9DSources-1.pdf}{http://}\\
\href{http://d279m997dpfwgl.cloudfront.net/wp/2016/11/Resource-False-Misleading-Clickbaity-and-Satirical-\%E2\%80\%9CNews\%E2\%80\%9DSources-1.pdf}{d279m997dpfwgl.cloudfront.net/wp/2016/11/Resource-False-Misleading-Clickbaity-and-}\\
\href{http://d279m997dpfwgl.cloudfront.net/wp/2016/11/Resource-False-Misleading-Clickbaity-and-Satirical-\%E2\%80\%9CNews\%E2\%80\%9DSources-1.pdf}{Satirical-\%E2\%80\%9CNews\%E2\%80\%9DSources-1.pdf}.
Accessed September 5, 2018.

Zubrzycki J (2017) More states take on media literacy in schools.
\emph{Education Week}. Available at \href{http://blogs.edweek.org/edweek/curriculum/2017/07/media_literacy_laws.html}{http://blogs.edweek.org/edweek/curriculum/2017/07/media\_literacy\_laws.html}.
Accessed September 3, 2018.

\parindent=2em
\leftskip=0em

\pagebreak
\begin{center}
\textbf{\Large Appendix}
\end{center}
\setcounter{equation}{0}
\setcounter{figure}{0}
\setcounter{table}{0}
\setcounter{section}{0}
\setcounter{footnote}{0}
\setcounter{page}{1}
\makeatletter
\setcounter{table}{0}
\renewcommand{\tablename}{Appendix Table}
\setcounter{figure}{0}
\renewcommand{\figurename}{Appendix Figure}

\section{Actions Against Fake News}

Appendix table \ref{tab:Actions-fb} lists Facebook's actions to reduce
the diffusion of fake news since the 2016 U.S. election, while Appendix
table \ref{tab:Actions-tw} lists Twitter's actions. All announcements
are taken from the platforms' official websites.\footnote{Facebook: \href{https://newsroom.fb.com/news/}{https://newsroom.fb.com/news/};
Twitter: \href{https://blog.twitter.com/official/en_us.html}{https://blog.twitter.com/official/en\_us.html}.}

\section{Data}

We combine five lists of fake news sites created by fact-checking
organizations or research studies to form our sample of fake news
sites. The union of these lists contains 673 unique sites. Among them,
103 have no data available from BuzzSumo. Thus, our final list includes
570 unique sites. Appendix table S3 presents the 50 largest sites
in the list in terms of total Facebook engagements plus Twitter shares
from January 2015 to July 2018.

We also collect three other categories of sites: major news sites,
small news sites, and business and culture sites covering arts, business,
health, recreation, and sports. Appendix table S4 presents these lists.

\section{Robustness Checks}

As discussed in the manuscript, a key concern is that our list of
fake news sites may suffer from sample selection bias. To mitigate
the concerns, we consider alternative sets of fake news sites as robustness
checks.

\subsection{Lists of Fake News Sites}

Our five different lists each have different inclusion criteria for
defining a fake news site, and one might disagree with a particular
list's approach. We thus carry out two sets of robustness checks.
First, in Appendix figure \ref{fig:multiple-lists}, we focus on sites
that are identified as fake news sites by at least two or three lists
instead of one, leaving 116 and 19 sites, respectively. Second, Appendix
figure \ref{fig:excluding-lists} replicates the results using sites
assembled from any four out of the five original lists. By doing this,
we exclude sites that are only identified by one particular list.
The downward trend in the ratio of Facebook engagements to Twitter
shares since the beginning of 2017 is invariant to including only
sites identified on multiple lists and to excluding any single list.

\subsection{Time Coverage}

It is possible that the original lists of fake news sites primarily
include sites that were popular on Facebook before the 2016 election,
and this sample selection combined with the rapid entry and exit of
small sites focused on fake stories could generate a spurious downward
trend in both the absolute number of Facebook engagements of fake
news and the ratio of Facebook engagements to Twitter shares. In Appendix
figure \ref{fig:time-coverage}, we look at sites that were active
during different periods. In Panel A, B, and C, respectively, we focus
on sites that started active operation after November 2016, sites
that were still in active operation as of the end of the sample in
July 2018, and sites that were in active operation from August 2015
to July 2018. (Active operation is defined to be a global traffic
rank reported by Alexa of at least one million.) The downward trend
in the ratio of Facebook engagements to Twitter shares since the beginning
of 2017 remains consistent across these samples.

\subsection{Number of Interactions}

Interactions on social media vary substantially across sites in our
list. A natural concern might be that the sums of Facebook engagements
and Twitter shares could be driven by a small number of outliers.
In Appendix figure \ref{fig:number-of-interactions} Panel A, we exclude
five largest sites in terms of total Facebook engagements plus Twitter
shares in our sample period. The trend survives the exclusion of potential
outliers. In Panel B and C, we divide all sites into deciles and look
at sites in the first decile and sites in the bottom nine deciles
separately. The downward trend in the Facebook/Twitter ratio is observed
for both large and small sites.

\subsection{Likelihood to Publish Misinformation}

Grinberg et al. (2018) provide three lists of sites which they deem
to have different likelihoods to publish misinformation. ``Black''
domains are reported to publish entirely fabricated stories. The black
list is constructed from pre-existing lists of fake news constructed
by academic work and professional fact-checkers such as PolitiFact,
FactCheck, and BuzzFeed. ``Red'' and ``orange'' domains are identified
by Snopes as sources of fake news or questionable claims and classified
by their levels of perceived likelihood to publish misinformation:
stories from red domains have an ``extremely high'' likelihood of
containing misinformation, and stories from orange domains a ``high''
likelihood. In Appendix figure \ref{fig:likelihood}, we look at these
lists separately. There are some differences across these lists. The
downward trend of Facebook engagements appears only for black and
red domains but not for orange domains. The time when the Facebook/Twitter
ratio started to fall is also different. For black domains, the ratio
dropped sharply in mid-2016 and all of 2017. For red and orange domains,
however, the decline primarily occurred in 2016. These patterns would
be consistent with black and to a lesser extent red domains being
the primary target of the changes Facebook made to its platform following
the election.

\subsection{Sites Focusing on Political News}

Fake news is often political in nature, and it is possible that the
Twitter user base is more consistently politically engaged than the
(much larger) Facebook user base. If Facebook users' interest in political
stories is cyclical, rising with major presidential elections and
falling after, this could generate a drop in fake news diffusion on
Facebook after the 2016 election that might not be mirrored on Twitter.
Thus, the declining Facebook/Twitter ratio beginning in 2017 could
be generated by changes in demand for fake news, not changes in supply
or efforts by Facebook.

\begin{figure}[H]
{\footnotesize{}\caption{Robustness Checks of Sites Focusing on Political News\label{fig:political}}
}{\footnotesize\par}

{\footnotesize{}\medskip{}
}{\footnotesize\par}
\begin{centering}
{\footnotesize{}}%
\begin{tabular}{c}
{\footnotesize{}\includegraphics[scale=0.65]{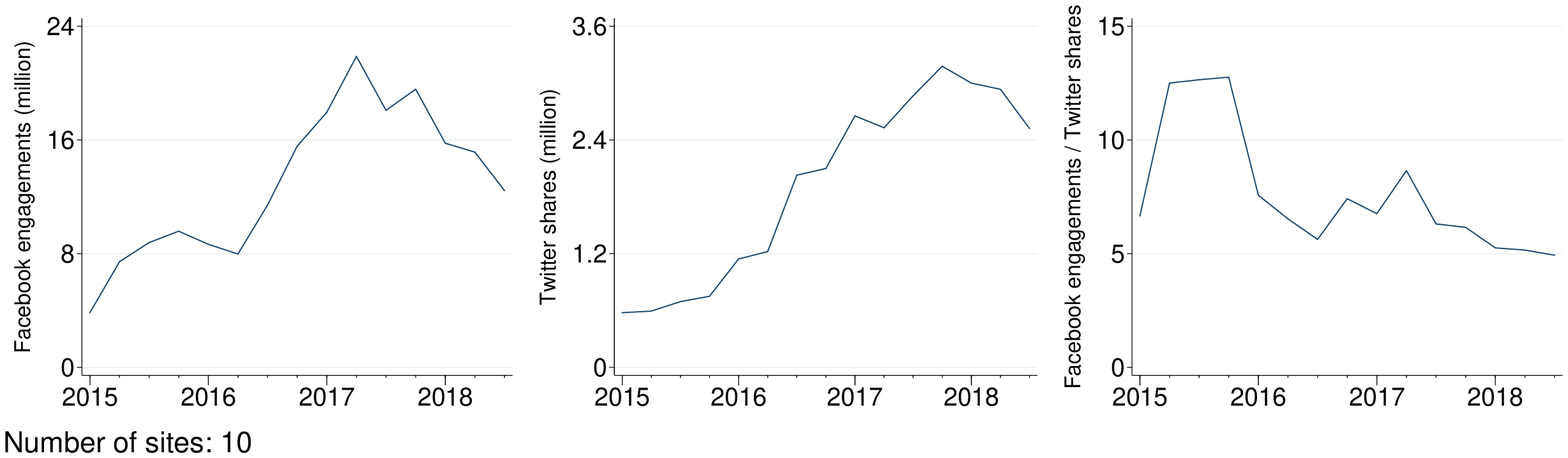}}\tabularnewline
\end{tabular}{\footnotesize\par}
\par\end{centering}
\medskip{}

\emph{\footnotesize{}Notes: }{\footnotesize{}This figure plots monthly
Facebook engagements, Twitter shares, and the ratio of Facebook engagements
over Twitter shares averaged by quarter of sites mostly focusing on
political news. The sites include politico.com, thehill.com, brookings.edu,
aei.org, c-span.org, realclearpolitics.com, donaldjtrump.com, hillaryclinton.com,
democrats.org, and gop.com.}{\footnotesize\par}
\end{figure}

If this explanation is true, one would also expect to see a decline
in the diffusion of articles from major political websites on Facebook,
but not on Twitter. To test this, Appendix figure \ref{fig:political}
presents Facebook engagements, Twitter shares, and their ratio for
a list of ten (non-fake) political sites of five types: (i) sites
mostly focusing on political news (Politico and The Hill); (ii) major
parties and politicians (donaldjtrump.com, hillaryclinton.com, democrats.org,
and gop.com); (iii) think tanks (Brookings and AEI); (iv) CSPAN; and
(v) a mainstream political blog (Real Clear Politics). There is a
decline in the Facebook/Twitter ratio for these sites, but it mainly
occured in late-2015, well before the election.

\parindent=2em
\leftskip=2em

\newpage{}

\begin{center}
\noindent\begin{minipage}[t]{1\columnwidth}%
\begin{table}[H]
\caption{Facebook's Actions to Fight Against Fake News\label{tab:Actions-fb}}

\medskip{}

\centering{}%
\begin{tabular}{l>{\raggedright}m{0.78\textwidth}}
\hline
\textbf{Date} & \textbf{Actions}\tabularnewline
\hline
Dec 15, 2016 & Announced four updates to address hoaxes and fake news: make reporting
easier for users; flag stories as \textquotedblleft Disputed\textquotedblright{}
with fact-checking organizations and warn people before they share;
incorporate signals of misleading articles into rankings; and disrupt
financial incentives for spammers.\tablefootnote{2.\href{https://newsroom.fb.com/news/2016/12/news-feed-fyi-addressing-hoaxes-and-fake-news/}{Addressing Hoaxes and Fake News.}}\tabularnewline
\hline
Apr 6, 2017 & Described three areas where it is working to fight the spead of false
news: disrupt economic incentives; build new products to curb the
spread of false news; and help people make more informed decisions.\tablefootnote{3.\href{https://newsroom.fb.com/news/2017/04/working-to-stop-misinformation-and-false-news/}{Working to Stop Misinformation and False News.}}\tabularnewline
\hline
Apr 25, 2017 & Tested \textquotedblleft Related Articles\textquotedblright , an improved
feature that presents users a cluster of additional articles on the
same topic when they come across popular links, including potantial
fake news articles, to provide people easier access to additional
information, including articles by third-party fact checkers.\tablefootnote{4.\href{https://newsroom.fb.com/news/2017/04/news-feed-fyi-new-test-with-related-articles/}{New Test With Related Articles.}}\tabularnewline
\hline
Aug 8, 2017 & Announced it would address cloaking so people see more authentic posts.\tablefootnote{5.\href{https://newsroom.fb.com/news/2017/08/news-feed-fyi-addressing-cloaking-so-people-see-more-authentic-posts/}{Addressing Cloaking So People See More Authentic Posts.}}\tabularnewline
\hline
Aug 28, 2017 & Announced it would block ads from pages that repeatedly share false
news.\tablefootnote{6.\href{https://newsroom.fb.com/news/2017/08/blocking-ads-from-pages-that-repeatedly-share-false-news/}{Blocking Ads From Pages that Repeatedly Share False News.}}\tabularnewline
\hline
Dec 20, 2017 & Annouced two changes to fight against false news: replace \textquotedblleft Disputed\textquotedblright{}
flags with \textquotedblleft Related Articles\textquotedblright{}
to give people more context; and start an initiative to better understand
how people decide whether information is accurate.\tablefootnote{7.\href{https://newsroom.fb.com/news/2017/12/news-feed-fyi-updates-in-our-fight-against-misinformation/}{Replacing Disputed Flags With Related Articles.}}\tabularnewline
\hline
Jan 11, 2018 & Prioritized posts from friends and family over public content.\tablefootnote{8.\href{https://newsroom.fb.com/news/2018/01/news-feed-fyi-bringing-people-closer-together/}{Bringing People Closer Together.}}\tabularnewline
\hline
Jan 19, 2018 & Prioritized news from publications rated as trustworthy by the community.\tablefootnote{9.\href{https://newsroom.fb.com/news/2018/01/trusted-sources/}{Helping Ensure News on Facebook Is From Trusted Sources.}}\tabularnewline
\hline
Jan 29, 2018 & Prioritized news relevant to people's local community.\tablefootnote{10.\href{https://newsroom.fb.com/news/2018/01/news-feed-fyi-local-news/}{More Local News on Facebook.}}\tabularnewline
\hline
May 23, 2018 & Described three parts of their strategies to stop misinformation:
remove accounts and content that violate community standards or ad
policies; reduce the distribution of false news and inauthentic content;
and inform people by giving them more context on the posts they see.\tablefootnote{11.\href{https://newsroom.fb.com/news/2018/05/hard-questions-false-news/}{Hard Questions: What\textquoteright s Facebook\textquoteright s Strategy for Stopping False News?}}\tabularnewline
\hline
June 14, 2018 & Detailed how its fact-checking program works.\tablefootnote{12.\href{https://newsroom.fb.com/news/2018/06/hard-questions-fact-checking/}{Hard Questions: How Is Facebook\textquoteright s Fact-Checking Program Working?}}\tabularnewline
\hline
June 21, 2018 & Announced five updates to fight false news: expand fact-checking programs
to new countries; test fact-checking on photos and videos; use new
techniques in fact-checking including identifying duplicates and using
\textquotedblleft Claim Review\textquotedblright ; take action against
repeat offenders; and improve measurement and transparency by partnering
with academics.\tablefootnote{13.\href{https://newsroom.fb.com/news/2018/06/increasing-our-efforts-to-fight-false-news/}{Increasing Our Efforts to Fight False News.}}\tabularnewline
\hline
\end{tabular}
\end{table}
\end{minipage}
\par\end{center}

\newpage{}
\begin{center}
\noindent\begin{minipage}[t]{1\columnwidth}%
\begin{table}[H]
\caption{Twitter's Actions to Fight Against Fake News\label{tab:Actions-tw}}

\medskip{}

\centering{}%
\begin{tabular}{l>{\raggedright}m{0.78\textwidth}}
\hline
\textbf{Date} & \textbf{Actions}\tabularnewline
\hline
June 14, 2017 & Described the phenomenon of fake news and bots and the approaches
it used, including surfacing the highest quality and most relevant
content and context first, expanding the team and resources, building
new tools and processes, and detecting spammy behaviors at source.
\tablefootnote{14.\href{https://blog.twitter.com/official/en_us/topics/company/2017/Our-Approach-Bots-Misinformation.html}{Our Approach to Bots \& Misinformation.}}\tabularnewline
\hline
June 29, 2017 & (Not officially announced) Tested a feature that would let users flag
tweets that contain misleading, false, or harmful information.\tablefootnote{15.\href{https://www.washingtonpost.com/news/the-switch/wp/2017/06/29/twitter-is-looking-for-ways-to-let-users-flag-fake-news/?utm_term=.9434cfc455c5}{Twitter is looking for ways to let users flag fake news, offensive content.}}\tabularnewline
\hline
Sept 28, 2017 & Shared information on its knowledge about how malicious bots and misinformation
networks on Twitter may have been used in the 2016 U.S. Presidential
elections and its work to fight both malicious bots and misinformation.\tablefootnote{16.\href{https://blog.twitter.com/official/en_us/topics/company/2017/Update-Russian-Interference-in-2016--Election-Bots-and-Misinformation.html}{Update: Russian Interference in 2016 US Election, Bots, \& Misinformation.}}\tabularnewline
\hline
Oct 24, 2017 & Announced steps to dramatically increase the transparency for all
ads.\tablefootnote{17.\href{https://blog.twitter.com/official/en_us/topics/product/2017/New-Transparency-For-Ads-on-Twitter.html}{New Transparency For Ads on Twitter.}}\tabularnewline
\hline
July 11, 2018 & Announced it removed fake accounts.\tablefootnote{18.\href{https://blog.twitter.com/official/en_us/topics/company/2018/Confidence-in-Follower-Counts.html}{Confidence in follower counts.}}\tabularnewline
\hline
\end{tabular}
\end{table}
\end{minipage}
\par\end{center}

\newpage{}
\begin{center}
Appendix Table 3: 50 Largest Fake News Sites\label{tab:fake}
\par\end{center}

\noindent \begin{center}
\begin{longtable}{lcccccccccc}
\hline
\hline
\multirow{2}{*}{\textbf{\scriptsize{}Site}} & \multicolumn{7}{c}{\textbf{\scriptsize{}Source}} & \multicolumn{1}{c}{\textbf{\scriptsize{}Created}} & \textbf{\scriptsize{}Still} & \textbf{\scriptsize{}Last}\tabularnewline
\cline{2-8}
 & \textbf{\scriptsize{}G-B} & \textbf{\scriptsize{}G-R} & \textbf{\scriptsize{}G-O} & \textbf{\scriptsize{}PF} & \textbf{\scriptsize{}BF} & \textbf{\scriptsize{}GNR} & \textbf{\scriptsize{}FC} & \textbf{\scriptsize{} Post-Election} & \textbf{\scriptsize{}Active} & \textbf{\scriptsize{}Long}\tabularnewline
\hline
\endhead
{\scriptsize{}indiatimes.com} & {\scriptsize{}0} & {\scriptsize{}0} & {\scriptsize{}0} & {\scriptsize{}1} & {\scriptsize{}0} & {\scriptsize{}0} & {\scriptsize{}0} & {\scriptsize{}0} & {\scriptsize{}1} & {\scriptsize{}1}\tabularnewline
\hline
{\scriptsize{}dailywire.com} & {\scriptsize{}0} & {\scriptsize{}0} & {\scriptsize{}1} & {\scriptsize{}0} & {\scriptsize{}0} & {\scriptsize{}1} & {\scriptsize{}0} & {\scriptsize{}0} & {\scriptsize{}1} & {\scriptsize{}0}\tabularnewline
\hline
{\scriptsize{}ijr.com} & {\scriptsize{}0} & {\scriptsize{}0} & {\scriptsize{}0} & {\scriptsize{}0} & {\scriptsize{}0} & {\scriptsize{}1} & {\scriptsize{}0} & {\scriptsize{}0} & {\scriptsize{}1} & {\scriptsize{}0}\tabularnewline
\hline
{\scriptsize{}dailycaller.com} & {\scriptsize{}0} & {\scriptsize{}0} & {\scriptsize{}1} & {\scriptsize{}0} & {\scriptsize{}0} & {\scriptsize{}0} & {\scriptsize{}0} & {\scriptsize{}0} & {\scriptsize{}1} & {\scriptsize{}1}\tabularnewline
\hline
{\scriptsize{}occupydemocrats.com} & {\scriptsize{}1} & {\scriptsize{}0} & {\scriptsize{}0} & {\scriptsize{}0} & {\scriptsize{}0} & {\scriptsize{}0} & {\scriptsize{}0} & {\scriptsize{}1} & {\scriptsize{}1} & {\scriptsize{}0}\tabularnewline
\hline
{\scriptsize{}express.co.uk} & {\scriptsize{}0} & {\scriptsize{}0} & {\scriptsize{}1} & {\scriptsize{}0} & {\scriptsize{}0} & {\scriptsize{}0} & {\scriptsize{}0} & {\scriptsize{}0} & {\scriptsize{}1} & {\scriptsize{}1}\tabularnewline
\hline
{\scriptsize{}redstatewatcher.com} & {\scriptsize{}1} & {\scriptsize{}0} & {\scriptsize{}0} & {\scriptsize{}0} & {\scriptsize{}0} & {\scriptsize{}1} & {\scriptsize{}0} & {\scriptsize{}0} & {\scriptsize{}1} & {\scriptsize{}0}\tabularnewline
\hline
{\scriptsize{}thepoliticalinsider.com} & {\scriptsize{}0} & {\scriptsize{}0} & {\scriptsize{}0} & {\scriptsize{}1} & {\scriptsize{}0} & {\scriptsize{}0} & {\scriptsize{}0} & {\scriptsize{}0} & {\scriptsize{}1} & {\scriptsize{}1}\tabularnewline
\hline
{\scriptsize{}thefederalistpapers.org} & {\scriptsize{}0} & {\scriptsize{}0} & {\scriptsize{}1} & {\scriptsize{}0} & {\scriptsize{}0} & {\scriptsize{}0} & {\scriptsize{}0} & {\scriptsize{}0} & {\scriptsize{}1} & {\scriptsize{}1}\tabularnewline
\hline
{\scriptsize{}truthfeed.com} & {\scriptsize{}0} & {\scriptsize{}1} & {\scriptsize{}0} & {\scriptsize{}0} & {\scriptsize{}0} & {\scriptsize{}1} & {\scriptsize{}0} & {\scriptsize{}0} & {\scriptsize{}1} & {\scriptsize{}0}\tabularnewline
\hline
{\scriptsize{}bipartisanreport.com} & {\scriptsize{}0} & {\scriptsize{}1} & {\scriptsize{}0} & {\scriptsize{}0} & {\scriptsize{}0} & {\scriptsize{}1} & {\scriptsize{}0} & {\scriptsize{}0} & {\scriptsize{}1} & {\scriptsize{}0}\tabularnewline
\hline
{\scriptsize{}rightwingnews.com} & {\scriptsize{}0} & {\scriptsize{}0} & {\scriptsize{}1} & {\scriptsize{}0} & {\scriptsize{}0} & {\scriptsize{}0} & {\scriptsize{}0} & {\scriptsize{}0} & {\scriptsize{}1} & {\scriptsize{}1}\tabularnewline
\hline
{\scriptsize{}qpolitical.com} & {\scriptsize{}0} & {\scriptsize{}0} & {\scriptsize{}1} & {\scriptsize{}0} & {\scriptsize{}0} & {\scriptsize{}0} & {\scriptsize{}0} & {\scriptsize{}0} & {\scriptsize{}0} & {\scriptsize{}0}\tabularnewline
\hline
{\scriptsize{}madworldnews.com} & {\scriptsize{}1} & {\scriptsize{}0} & {\scriptsize{}0} & {\scriptsize{}1} & {\scriptsize{}0} & {\scriptsize{}0} & {\scriptsize{}0} & {\scriptsize{}0} & {\scriptsize{}1} & {\scriptsize{}1}\tabularnewline
\hline
{\scriptsize{}yournewswire.com} & {\scriptsize{}1} & {\scriptsize{}0} & {\scriptsize{}0} & {\scriptsize{}1} & {\scriptsize{}1} & {\scriptsize{}0} & {\scriptsize{}1} & {\scriptsize{}0} & {\scriptsize{}1} & {\scriptsize{}1}\tabularnewline
\hline
{\scriptsize{}uschronicle.com} & {\scriptsize{}0} & {\scriptsize{}0} & {\scriptsize{}1} & {\scriptsize{}0} & {\scriptsize{}0} & {\scriptsize{}0} & {\scriptsize{}0} & {\scriptsize{}0} & {\scriptsize{}1} & {\scriptsize{}0}\tabularnewline
\hline
{\scriptsize{}louderwithcrowder.com} & {\scriptsize{}0} & {\scriptsize{}1} & {\scriptsize{}0} & {\scriptsize{}0} & {\scriptsize{}0} & {\scriptsize{}0} & {\scriptsize{}0} & {\scriptsize{}0} & {\scriptsize{}1} & {\scriptsize{}1}\tabularnewline
\hline
{\scriptsize{}jewsnews.co.il} & {\scriptsize{}0} & {\scriptsize{}0} & {\scriptsize{}0} & {\scriptsize{}1} & {\scriptsize{}0} & {\scriptsize{}0} & {\scriptsize{}0} & {\scriptsize{}0} & {\scriptsize{}1} & {\scriptsize{}1}\tabularnewline
\hline
{\scriptsize{}100percentfedup.com} & {\scriptsize{}0} & {\scriptsize{}1} & {\scriptsize{}0} & {\scriptsize{}0} & {\scriptsize{}0} & {\scriptsize{}0} & {\scriptsize{}0} & {\scriptsize{}0} & {\scriptsize{}1} & {\scriptsize{}1}\tabularnewline
\hline
{\scriptsize{}angrypatriotmovement.com} & {\scriptsize{}1} & {\scriptsize{}0} & {\scriptsize{}0} & {\scriptsize{}1} & {\scriptsize{}0} & {\scriptsize{}1} & {\scriptsize{}0} & {\scriptsize{}0} & {\scriptsize{}1} & {\scriptsize{}0}\tabularnewline
\hline
{\scriptsize{}anonhq.com} & {\scriptsize{}0} & {\scriptsize{}1} & {\scriptsize{}0} & {\scriptsize{}0} & {\scriptsize{}0} & {\scriptsize{}0} & {\scriptsize{}0} & {\scriptsize{}0} & {\scriptsize{}1} & {\scriptsize{}1}\tabularnewline
\hline
{\scriptsize{}inquisitr.com} & {\scriptsize{}0} & {\scriptsize{}0} & {\scriptsize{}1} & {\scriptsize{}0} & {\scriptsize{}0} & {\scriptsize{}0} & {\scriptsize{}0} & {\scriptsize{}0} & {\scriptsize{}1} & {\scriptsize{}1}\tabularnewline
\hline
{\scriptsize{}yesimright.com} & {\scriptsize{}1} & {\scriptsize{}0} & {\scriptsize{}0} & {\scriptsize{}0} & {\scriptsize{}0} & {\scriptsize{}1} & {\scriptsize{}0} & {\scriptsize{}0} & {\scriptsize{}0} & {\scriptsize{}0}\tabularnewline
\hline
{\scriptsize{}worldtruth.tv} & {\scriptsize{}0} & {\scriptsize{}1} & {\scriptsize{}0} & {\scriptsize{}1} & {\scriptsize{}0} & {\scriptsize{}0} & {\scriptsize{}0} & {\scriptsize{}0} & {\scriptsize{}1} & {\scriptsize{}1}\tabularnewline
\hline
{\scriptsize{}collective-evolution.com} & {\scriptsize{}0} & {\scriptsize{}1} & {\scriptsize{}0} & {\scriptsize{}0} & {\scriptsize{}0} & {\scriptsize{}0} & {\scriptsize{}0} & {\scriptsize{}0} & {\scriptsize{}1} & {\scriptsize{}1}\tabularnewline
\hline
{\scriptsize{}ilovemyfreedom.org} & {\scriptsize{}1} & {\scriptsize{}0} & {\scriptsize{}0} & {\scriptsize{}0} & {\scriptsize{}0} & {\scriptsize{}1} & {\scriptsize{}0} & {\scriptsize{}0} & {\scriptsize{}1} & {\scriptsize{}0}\tabularnewline
\hline
{\scriptsize{}tribunist.com} & {\scriptsize{}0} & {\scriptsize{}0} & {\scriptsize{}1} & {\scriptsize{}0} & {\scriptsize{}0} & {\scriptsize{}0} & {\scriptsize{}0} & {\scriptsize{}0} & {\scriptsize{}1} & {\scriptsize{}0}\tabularnewline
\hline
{\scriptsize{}clashdaily.com} & {\scriptsize{}1} & {\scriptsize{}0} & {\scriptsize{}0} & {\scriptsize{}1} & {\scriptsize{}0} & {\scriptsize{}0} & {\scriptsize{}0} & {\scriptsize{}0} & {\scriptsize{}1} & {\scriptsize{}1}\tabularnewline
\hline
{\scriptsize{}naturalnews.com} & {\scriptsize{}0} & {\scriptsize{}1} & {\scriptsize{}0} & {\scriptsize{}0} & {\scriptsize{}0} & {\scriptsize{}0} & {\scriptsize{}0} & {\scriptsize{}0} & {\scriptsize{}1} & {\scriptsize{}1}\tabularnewline
\hline
{\scriptsize{}joeforamerica.com} & {\scriptsize{}0} & {\scriptsize{}0} & {\scriptsize{}1} & {\scriptsize{}0} & {\scriptsize{}0} & {\scriptsize{}0} & {\scriptsize{}0} & {\scriptsize{}0} & {\scriptsize{}1} & {\scriptsize{}1}\tabularnewline
\hline
{\scriptsize{}conservativedailypost.com} & {\scriptsize{}1} & {\scriptsize{}0} & {\scriptsize{}0} & {\scriptsize{}1} & {\scriptsize{}0} & {\scriptsize{}1} & {\scriptsize{}0} & {\scriptsize{}0} & {\scriptsize{}1} & {\scriptsize{}0}\tabularnewline
\hline
{\scriptsize{}worldnewsdailyreport.com} & {\scriptsize{}1} & {\scriptsize{}0} & {\scriptsize{}0} & {\scriptsize{}1} & {\scriptsize{}1} & {\scriptsize{}0} & {\scriptsize{}0} & {\scriptsize{}0} & {\scriptsize{}1} & {\scriptsize{}1}\tabularnewline
\hline
{\scriptsize{}trueactivist.com} & {\scriptsize{}0} & {\scriptsize{}0} & {\scriptsize{}1} & {\scriptsize{}0} & {\scriptsize{}0} & {\scriptsize{}0} & {\scriptsize{}0} & {\scriptsize{}0} & {\scriptsize{}1} & {\scriptsize{}1}\tabularnewline
\hline
{\scriptsize{}americasfreedomfighters.com} & {\scriptsize{}0} & {\scriptsize{}1} & {\scriptsize{}0} & {\scriptsize{}0} & {\scriptsize{}0} & {\scriptsize{}0} & {\scriptsize{}0} & {\scriptsize{}0} & {\scriptsize{}1} & {\scriptsize{}1}\tabularnewline
\hline
{\scriptsize{}conservative101.com} & {\scriptsize{}0} & {\scriptsize{}0} & {\scriptsize{}0} & {\scriptsize{}1} & {\scriptsize{}0} & {\scriptsize{}0} & {\scriptsize{}0} & {\scriptsize{}0} & {\scriptsize{}1} & {\scriptsize{}0}\tabularnewline
\hline
{\scriptsize{}usanewsflash.com} & {\scriptsize{}1} & {\scriptsize{}0} & {\scriptsize{}0} & {\scriptsize{}1} & {\scriptsize{}0} & {\scriptsize{}1} & {\scriptsize{}0} & {\scriptsize{}0} & {\scriptsize{}0} & {\scriptsize{}0}\tabularnewline
\hline
{\scriptsize{}babylonbee.com} & {\scriptsize{}0} & {\scriptsize{}0} & {\scriptsize{}0} & {\scriptsize{}1} & {\scriptsize{}0} & {\scriptsize{}0} & {\scriptsize{}0} & {\scriptsize{}0} & {\scriptsize{}1} & {\scriptsize{}0}\tabularnewline
\hline
{\scriptsize{}firstpost.com} & {\scriptsize{}0} & {\scriptsize{}0} & {\scriptsize{}0} & {\scriptsize{}1} & {\scriptsize{}0} & {\scriptsize{}0} & {\scriptsize{}0} & {\scriptsize{}0} & {\scriptsize{}1} & {\scriptsize{}1}\tabularnewline
\hline
{\scriptsize{}zerohedge.com} & {\scriptsize{}0} & {\scriptsize{}0} & {\scriptsize{}1} & {\scriptsize{}0} & {\scriptsize{}0} & {\scriptsize{}0} & {\scriptsize{}0} & {\scriptsize{}0} & {\scriptsize{}1} & {\scriptsize{}1}\tabularnewline
\hline
{\scriptsize{}teaparty.org} & {\scriptsize{}0} & {\scriptsize{}0} & {\scriptsize{}0} & {\scriptsize{}1} & {\scriptsize{}0} & {\scriptsize{}0} & {\scriptsize{}0} & {\scriptsize{}0} & {\scriptsize{}1} & {\scriptsize{}1}\tabularnewline
\hline
{\scriptsize{}palmerreport.com} & {\scriptsize{}0} & {\scriptsize{}0} & {\scriptsize{}1} & {\scriptsize{}0} & {\scriptsize{}0} & {\scriptsize{}0} & {\scriptsize{}0} & {\scriptsize{}1} & {\scriptsize{}1} & {\scriptsize{}0}\tabularnewline
\hline
{\scriptsize{}judicialwatch.org} & {\scriptsize{}0} & {\scriptsize{}1} & {\scriptsize{}0} & {\scriptsize{}0} & {\scriptsize{}0} & {\scriptsize{}0} & {\scriptsize{}0} & {\scriptsize{}0} & {\scriptsize{}1} & {\scriptsize{}1}\tabularnewline
\hline
{\scriptsize{}disclose.tv} & {\scriptsize{}1} & {\scriptsize{}0} & {\scriptsize{}0} & {\scriptsize{}1} & {\scriptsize{}0} & {\scriptsize{}0} & {\scriptsize{}0} & {\scriptsize{}0} & {\scriptsize{}1} & {\scriptsize{}1}\tabularnewline
\hline
{\scriptsize{}conservativepost.com} & {\scriptsize{}0} & {\scriptsize{}1} & {\scriptsize{}0} & {\scriptsize{}0} & {\scriptsize{}0} & {\scriptsize{}0} & {\scriptsize{}0} & {\scriptsize{}0} & {\scriptsize{}1} & {\scriptsize{}1}\tabularnewline
\hline
{\scriptsize{}thegatewaypundit.com} & {\scriptsize{}0} & {\scriptsize{}1} & {\scriptsize{}0} & {\scriptsize{}1} & {\scriptsize{}0} & {\scriptsize{}0} & {\scriptsize{}0} & {\scriptsize{}0} & {\scriptsize{}1} & {\scriptsize{}1}\tabularnewline
\hline
{\scriptsize{}infowars.com} & {\scriptsize{}0} & {\scriptsize{}1} & {\scriptsize{}0} & {\scriptsize{}0} & {\scriptsize{}0} & {\scriptsize{}0} & {\scriptsize{}0} & {\scriptsize{}0} & {\scriptsize{}1} & {\scriptsize{}1}\tabularnewline
\hline
{\scriptsize{}dailysnark.com} & {\scriptsize{}0} & {\scriptsize{}0} & {\scriptsize{}0} & {\scriptsize{}1} & {\scriptsize{}0} & {\scriptsize{}0} & {\scriptsize{}0} & {\scriptsize{}0} & {\scriptsize{}1} & {\scriptsize{}1}\tabularnewline
\hline
{\scriptsize{}postcard.news} & {\scriptsize{}0} & {\scriptsize{}0} & {\scriptsize{}0} & {\scriptsize{}1} & {\scriptsize{}0} & {\scriptsize{}0} & {\scriptsize{}0} & {\scriptsize{}0} & {\scriptsize{}1} & {\scriptsize{}0}\tabularnewline
\hline
{\scriptsize{}higherperspectives.com} & {\scriptsize{}0} & {\scriptsize{}0} & {\scriptsize{}0} & {\scriptsize{}1} & {\scriptsize{}0} & {\scriptsize{}0} & {\scriptsize{}0} & {\scriptsize{}0} & {\scriptsize{}1} & {\scriptsize{}0}\tabularnewline
\hline
{\scriptsize{}tmn.today} & {\scriptsize{}0} & {\scriptsize{}0} & {\scriptsize{}1} & {\scriptsize{}0} & {\scriptsize{}0} & {\scriptsize{}1} & {\scriptsize{}0} & {\scriptsize{}0} & {\scriptsize{}1} & {\scriptsize{}0}\tabularnewline
\hline
\multicolumn{11}{c}{{\scriptsize{}...}}\tabularnewline
\hline
\textbf{\scriptsize{}Total} & \textbf{\scriptsize{}382} & \textbf{\scriptsize{}61} & \textbf{\scriptsize{}47} & \textbf{\scriptsize{}325} & \textbf{\scriptsize{}223} & \textbf{\scriptsize{}92} & \textbf{\scriptsize{}61} & \textbf{\scriptsize{}308} & \textbf{\scriptsize{}287} & \textbf{\scriptsize{}82}\tabularnewline
\hline
\end{longtable}
\par\end{center}

\begin{onehalfspace}
\noindent \emph{\footnotesize{}Note}{\footnotesize{}s: This table
lists 50 largest fake news sites in terms of total Facebook engagements
plus Twitter shares from January 2015 to June 2018. The complete list
can be found \href{https://docs.google.com/spreadsheets/d/1OW8qg_PqOzaclNXeZMVJXfKhx86UzbuCa8jQEEuOqQQ/edit?usp=sharing}{here}.
Column 2-8 lists the fake news sites identified by five sources described
above, where a value of 1 indicates the site appears in the corresponding
source and 0 not. }\textbf{\footnotesize{}G-B}{\footnotesize{},}\textbf{\footnotesize{}
G-R}{\footnotesize{}, and }\textbf{\footnotesize{}G-O }{\footnotesize{}represent
the black domains, red domains, and orange domains in Grinberg et
al. (2018). }\textbf{\footnotesize{}PF}{\footnotesize{} represents
PolitiFact. }\textbf{\footnotesize{}BF}{\footnotesize{} represents
BuzzFeed.}\textbf{\footnotesize{} GNR}{\footnotesize{} represents
Guess et al. (2018). }\textbf{\footnotesize{}FC }{\footnotesize{}represents
FactCheck. The last three columns list sites that started active operation
after the election in November 2016, sites that were in active operation
in July 2018, and sites that were in active operation during the whole
sample period from August 2015 to July 2018. A site is defined as
being in active operation if it is tracked by Alexa with a global
rank higher than one million in terms of total traffic.}{\footnotesize\par}
\end{onehalfspace}

\newpage{}
\begin{center}
Appendix Table 4: Lists of Sites in Each Category\label{tab:other}
\par\end{center}

\begin{center}
\begin{longtable}{llll}
\hline
\hline
\textbf{\scriptsize{}Category} & \multicolumn{3}{c}{\textbf{\scriptsize{}Site}}\tabularnewline
\hline
\endhead
\multirow{13}{*}{\textbf{\emph{\scriptsize{}Major News Sites}}} & {\scriptsize{}cnn.com } & {\scriptsize{}nytimes.com} & {\scriptsize{}theguardian.com}\tabularnewline
 & {\scriptsize{}washingtonpost.com} & {\scriptsize{}foxnews.com} & {\scriptsize{}huffingtonpost.com}\tabularnewline
 & {\scriptsize{}usatoday.com} & {\scriptsize{}wsj.com} & {\scriptsize{}cnbc.com}\tabularnewline
 & {\scriptsize{}reuters.com } & {\scriptsize{}time.com} & {\scriptsize{}nypost.com}\tabularnewline
 & {\scriptsize{}usnews.com} & {\scriptsize{}cbsnews.com} & {\scriptsize{}chron.com}\tabularnewline
 & {\scriptsize{}thehill.com} & {\scriptsize{}nbcnews.com} & {\scriptsize{}theatlantic.com}\tabularnewline
 & {\scriptsize{}latimes.com} & {\scriptsize{}abcnews.go.com} & {\scriptsize{}thedailybeast.com}\tabularnewline
 & {\scriptsize{}sfgate.com} & {\scriptsize{}newsweek.com} & {\scriptsize{}chicagotribune.com}\tabularnewline
 & {\scriptsize{}economist.com} & {\scriptsize{}theroot.com} & {\scriptsize{}voanews.com}\tabularnewline
 & {\scriptsize{}nj.com} & {\scriptsize{}miamiherald.com} & {\scriptsize{}mercurynews.com}\tabularnewline
 & {\scriptsize{}bostonglobe.com} & {\scriptsize{}seattletimes.com} & {\scriptsize{}oregonlive.com}\tabularnewline
 & {\scriptsize{}washingtontimes.com} & {\scriptsize{}azcentral.com} & {\scriptsize{}ajc.com}\tabularnewline
 & {\scriptsize{}philly.com} & {\scriptsize{}sacbee.com} & \tabularnewline
\hline
\multirow{24}{*}{\textbf{\emph{\scriptsize{}Small News Sites}}} & {\scriptsize{}aspentimes.com} & {\scriptsize{}bakersfield.com} & {\scriptsize{}bendbulletin.com}\tabularnewline
 & {\scriptsize{}bnd.com} & {\scriptsize{}broadcastingcable.com} & {\scriptsize{}charlestoncitypaper.com}\tabularnewline
 & {\scriptsize{}chicagomaroon.com} & {\scriptsize{}collegian.psu.edu} & {\scriptsize{}columbian.com}\tabularnewline
 & {\scriptsize{}dailynebraskan.com} & {\scriptsize{}dailynexus.com} & {\scriptsize{}dailynorthwestern.com}\tabularnewline
 & {\scriptsize{}dailypress.com} & {\scriptsize{}dailyprogress.com} & {\scriptsize{}dailytexanonline.com}\tabularnewline
 & {\scriptsize{}dailytrojan.com} & {\scriptsize{}dcourier.com} & {\scriptsize{}delcotimes.com}\tabularnewline
 & {\scriptsize{}durangoherald.com} & {\scriptsize{}fair.org} & {\scriptsize{}fredericksburg.com}\tabularnewline
 & {\scriptsize{}globegazette.com} & {\scriptsize{}greenvilleonline.com} & {\scriptsize{}greenwichtime.com}\tabularnewline
 & {\scriptsize{}havasunews.com} & {\scriptsize{}hcn.org} & {\scriptsize{}heraldnet.com}\tabularnewline
 & {\scriptsize{}heraldsun.com} & {\scriptsize{}heraldtimesonline.com} & {\scriptsize{}ibj.com}\tabularnewline
 & {\scriptsize{}independent.com} & {\scriptsize{}islandpacket.com} & {\scriptsize{}jou.ufl.edu}\tabularnewline
 & {\scriptsize{}journalism.org} & {\scriptsize{}journalismjobs.com} & {\scriptsize{}journaltimes.com}\tabularnewline
 & {\scriptsize{}kitv.com} & {\scriptsize{}knoxnews.com} & {\scriptsize{}lacrossetribune.com}\tabularnewline
 & {\scriptsize{}leadertelegram.com} & {\scriptsize{}macon.com} & {\scriptsize{}myrtlebeachonline.com}\tabularnewline
 & {\scriptsize{}naplesnews.com} & {\scriptsize{}nashvillescene.com} & {\scriptsize{}news.cornell.edu}\tabularnewline
 & {\scriptsize{}news.usc.edu} & {\scriptsize{}newseum.org} & {\scriptsize{}news-journalonline.com}\tabularnewline
 & {\scriptsize{}news-leader.com } & {\scriptsize{}newstimes.com} & {\scriptsize{}nwfdailynews.com}\tabularnewline
 & {\scriptsize{}pjstar.com} & {\scriptsize{}presstelegram.com} & {\scriptsize{}rapidcityjournal.com}\tabularnewline
 & {\scriptsize{}readingeagle.com} & {\scriptsize{}redandblack.com} & {\scriptsize{}rgj.com}\tabularnewline
 & {\scriptsize{}sacurrent.com} & {\scriptsize{}santacruzsentinel.com} & {\scriptsize{}santafenewmexican.com}\tabularnewline
 \hline
\multirow{6}{*}{\textbf{\emph{\scriptsize{}Small News Sites}}} 
 & {\scriptsize{}sgvtribune.com} & {\scriptsize{}signalscv.com} & {\scriptsize{}siouxcityjournal.com}\tabularnewline
 & {\scriptsize{}standard.net} & {\scriptsize{}stanforddaily.com} & {\scriptsize{}steynonline.com}\tabularnewline
 & {\scriptsize{}studlife.com} & {\scriptsize{}tallahassee.com} & {\scriptsize{}theday.com}\tabularnewline
 & {\scriptsize{}theeagle.com} & {\scriptsize{}theledger.com} & {\scriptsize{}timesleader.com}\tabularnewline
 & {\scriptsize{}ubm.com} & {\scriptsize{}vcstar.com} & {\scriptsize{}wacotrib.com}\tabularnewline
 & {\scriptsize{}wcfcourier.com} & {\scriptsize{}wvgazettemail.com} & {\scriptsize{}yakimaherald.com}\tabularnewline
\hline
\multirow{3}{*}{\textbf{\emph{\scriptsize{}Arts}}} & {\scriptsize{}imdb.com} & {\scriptsize{}ign.com} & {\scriptsize{}rottentomatoes.com}\tabularnewline
 & {\scriptsize{}ultimate-guitar.com} & {\scriptsize{}npr.org} & {\scriptsize{}vice.com}\tabularnewline
 & {\scriptsize{}tmz.com} & {\scriptsize{}pitchfork.com} & {\scriptsize{}wired.com}\tabularnewline
\hline
\multirow{3}{*}{\textbf{\emph{\scriptsize{}Business}}} & {\scriptsize{}forbes.com} & {\scriptsize{}shutterstock.com} & {\scriptsize{}businessinsider.com}\tabularnewline
 & {\scriptsize{}finance.yahoo.com} & {\scriptsize{}bloomberg.com} & {\scriptsize{}eventbrite.com}\tabularnewline
 & {\scriptsize{}fortune.com} & {\scriptsize{}adweek.com} & \tabularnewline
\hline
\multirow{3}{*}{\textbf{\emph{\scriptsize{}Health}}} & {\scriptsize{}webmd.com} & {\scriptsize{}psychologytoday.com} & {\scriptsize{}who.int}\tabularnewline
 & {\scriptsize{}apa.org} & {\scriptsize{}bmj.com} & {\scriptsize{}mercola.com}\tabularnewline
 & {\scriptsize{}menshealth.com} & {\scriptsize{}self.com} & {\scriptsize{}nejm.org}\tabularnewline
\hline
\multirow{3}{*}{\textbf{\emph{\scriptsize{}Recreation}}} & {\scriptsize{}9gag.com } & {\scriptsize{}jalopnik.com} & {\scriptsize{}timeout.com}\tabularnewline
 & {\scriptsize{}lonelyplanet.com} & {\scriptsize{}caranddriver.com} & {\scriptsize{}hollywoodreporter.com}\tabularnewline
 & {\scriptsize{}nationalgeographic.com} & {\scriptsize{}rd.com} & {\scriptsize{}topix.com}\tabularnewline
\hline
\multirow{7}{*}{\textbf{\emph{\scriptsize{}Sports}}} & {\scriptsize{}espn.com} & {\scriptsize{}cricbuzz.com} & {\scriptsize{}nba.com}\tabularnewline
 & {\scriptsize{}espncricinfo.com} & {\scriptsize{}sports.yahoo.com} & {\scriptsize{}bleacherreport.com}\tabularnewline
 & {\scriptsize{}nhl.com} & {\scriptsize{}cbssports.com} & {\scriptsize{}nfl.com}\tabularnewline
 & {\scriptsize{}iplt20.com} & {\scriptsize{}skysports.com} & {\scriptsize{}deadspin.com}\tabularnewline
 & {\scriptsize{}nbcsports.com} & {\scriptsize{}wwe.com} & {\scriptsize{}si.com}\tabularnewline
 & {\scriptsize{}sbnation.com} & {\scriptsize{}formula1.com} & {\scriptsize{}rivals.com}\tabularnewline
 & {\scriptsize{}foxsports.com} &  & \tabularnewline
\hline
\end{longtable}
\par\end{center}

\noindent \emph{\footnotesize{}Note}{\footnotesize{}s: This table
lists sites in the comparison groups. }\emph{\footnotesize{}Major
News Sites}{\footnotesize{} include 38 sites selected from the top
100 sites in Alexa's News category. }\emph{\footnotesize{}Small News
Sites}{\footnotesize{} include 78 sites selected from the sites ranking
401-500 in the News category. }\emph{\footnotesize{}Business and Culture
Sites}{\footnotesize{} include 54 sites selected from the top 50 sites
in each of the Arts, Business, Health, Recreation, and Sports categories.}\emph{\footnotesize{}
}{\footnotesize{}For each group, we omit from our sample government
websites, databases, sites that do not mainly produce news or similar
content, international sites whose audiences are primarily outside
the U.S., and sites that are included in our list of fake news sites.}{\footnotesize\par}

\noindent {\footnotesize{}\newpage}{\footnotesize\par}

\begin{figure}[H]
{\footnotesize{}\caption{Robustness Checks of Fake News Sites - Multiple Lists\label{fig:multiple-lists}}
}{\footnotesize\par}

{\footnotesize{}\medskip{}
}{\footnotesize\par}
\begin{centering}
{\footnotesize{}}%
\begin{tabular}{c}
\emph{Panel A: Sites Identified by At Least Two Lists}\tabularnewline
{\footnotesize{}\includegraphics[scale=0.65]{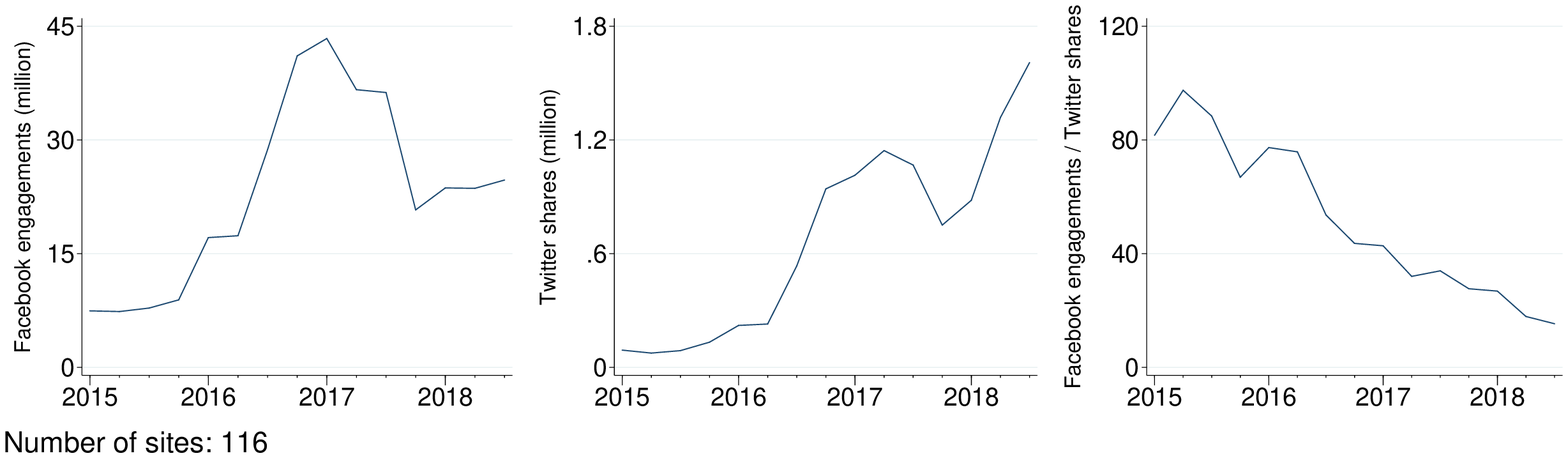}}\tabularnewline
\emph{Panel B: Sites Identified by At Least Three Lists}\tabularnewline
{\footnotesize{}\includegraphics[scale=0.65]{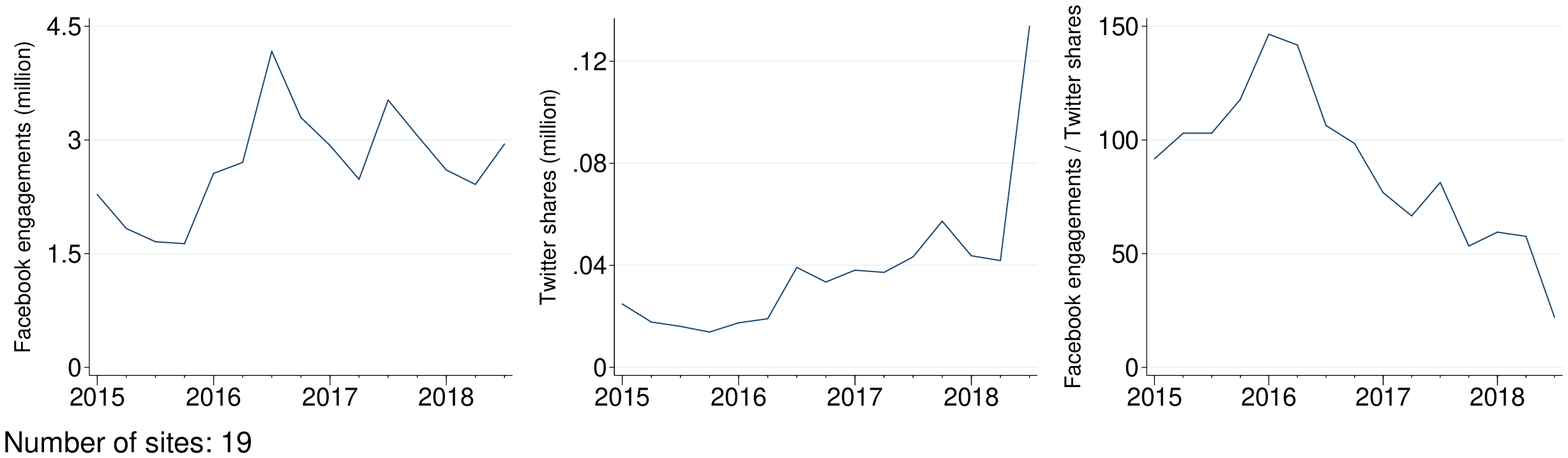}}\tabularnewline
\end{tabular}{\footnotesize\par}
\par\end{centering}
\medskip{}

\emph{\footnotesize{}Notes: }{\footnotesize{}This figure plots robustness
checks for the sample of fake news sites by looking at sites only
identified by multiple lists. Each panel plots monthly Facebook engagements,
Twitter shares, and the ratio of Facebook engagements over Twitter
shares averaged by quarter. Panel A includes sites identified by at
least two lists out of five. Panel B includes sites identified by
at least three lists. Grinberg et al.'s (2018) provide three types
of domains. The black domains derive from lists that we already use
(with the exception of nine sites, as PolitiFact and FactCheck updated
their lists at some point). We avoid double-counting black domains
when we count the number of lists that identify a fake news site.}{\footnotesize\par}
\end{figure}

\newpage{}

\begin{figure}[H]
{\footnotesize{}\caption{Robustness Checks of Fake News Sites - Excluding Lists\label{fig:excluding-lists}}
}{\footnotesize\par}

{\footnotesize{}\medskip{}
}{\footnotesize\par}
\centering{}{\footnotesize{}}%
\begin{tabular}{c}
\emph{Panel A: Excluding Sites Only Identified by Grinberg et al.
(2018)}\tabularnewline
{\footnotesize{}\includegraphics[scale=0.65]{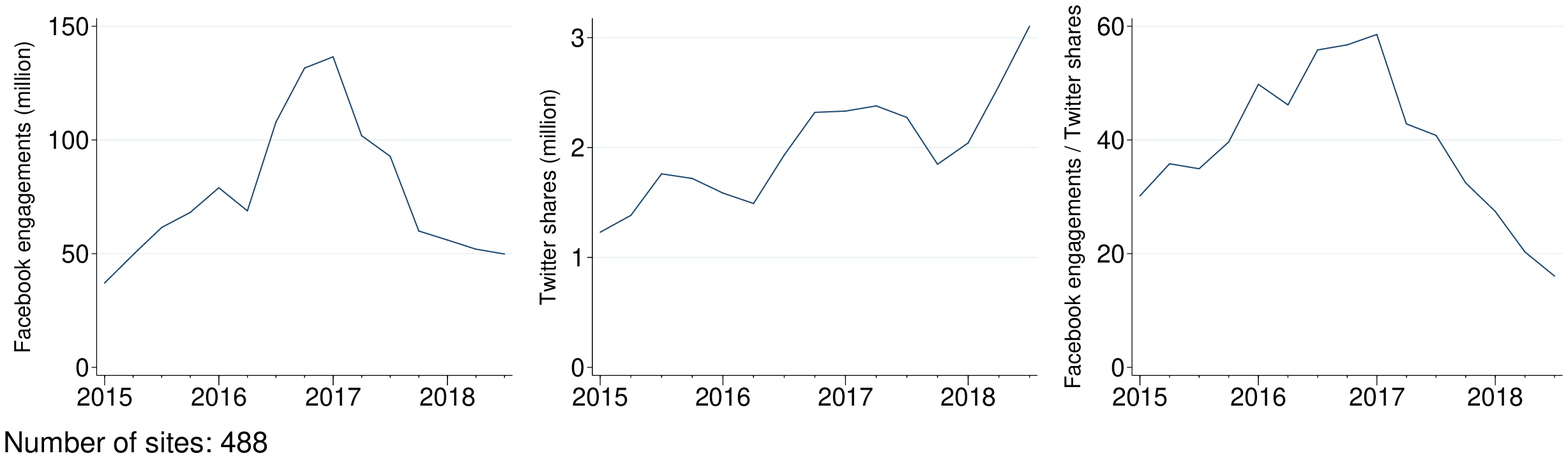}}\tabularnewline
\emph{Panel B: Excluding Sites Only Ideitifed by PolitiFact}\tabularnewline
{\footnotesize{}\includegraphics[scale=0.65]{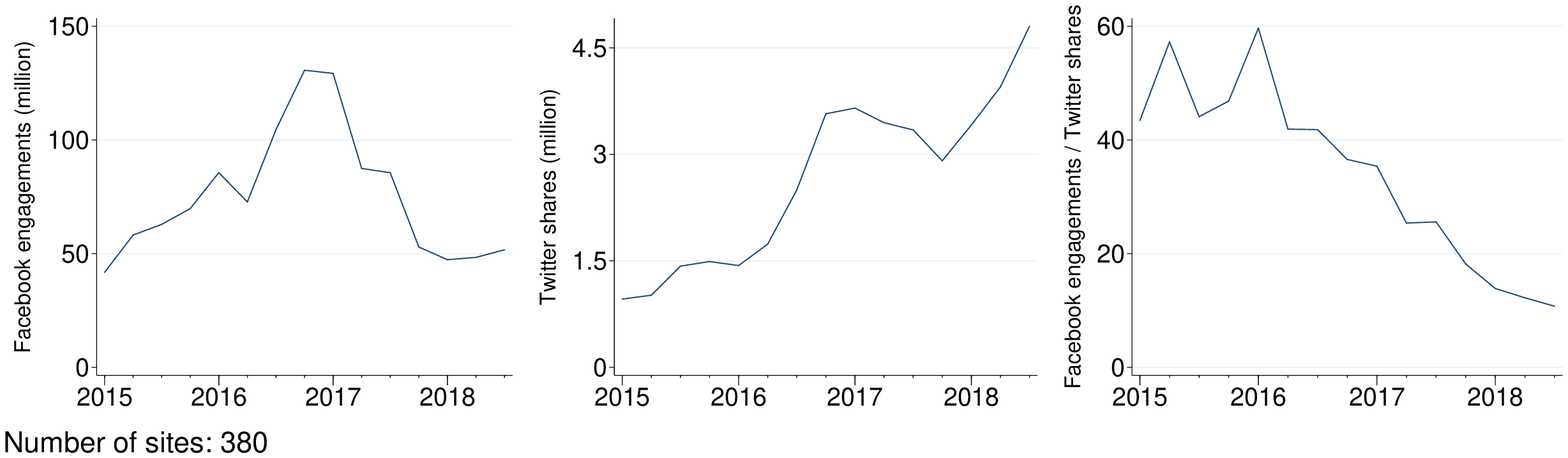}}\tabularnewline
\emph{Panel C: Excluding Sites Only Identified by BuzzFeed}\tabularnewline
{\footnotesize{}\includegraphics[scale=0.65]{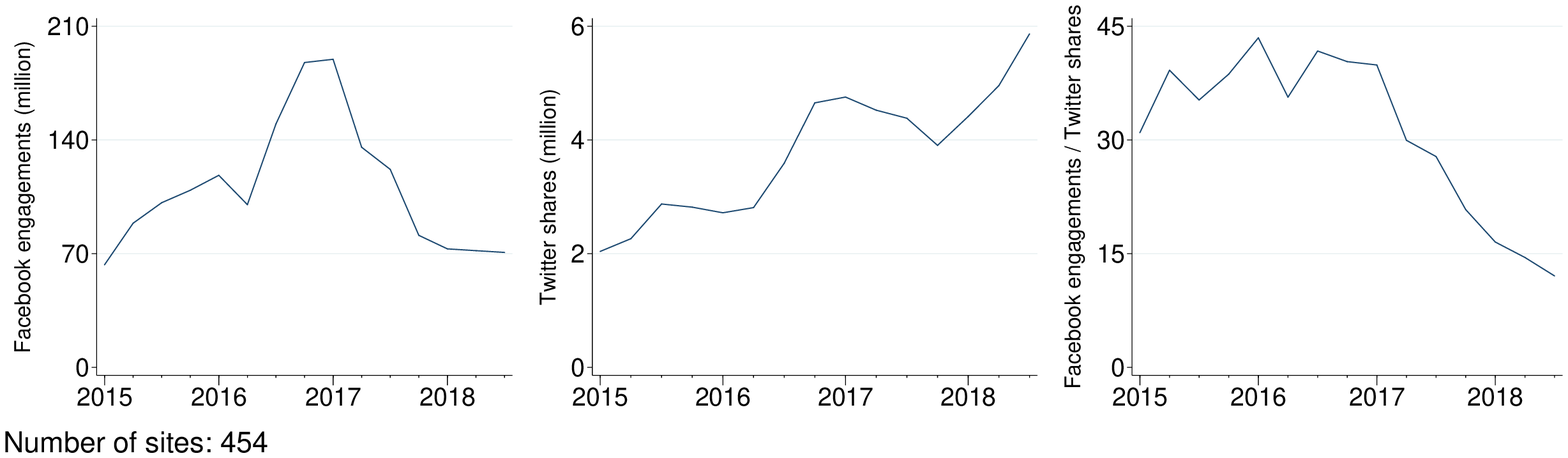}}\tabularnewline
\end{tabular}{\footnotesize\par}
\end{figure}

\begin{figure}[H]
\begin{centering}
{\footnotesize{}\ContinuedFloat}{\footnotesize\par}
\par\end{centering}
{\footnotesize{}\caption{Robustness Checks of Fake News Sites - Excluding Lists (\emph{continued})}
}{\footnotesize\par}

{\footnotesize{}\medskip{}
}{\footnotesize\par}
\begin{centering}
{\footnotesize{}}%
\begin{tabular}{c}
\emph{Panel D: Excluding Sites Only Ideitifed by Guess et al. (2018)}\tabularnewline
{\footnotesize{}\includegraphics[scale=0.65]{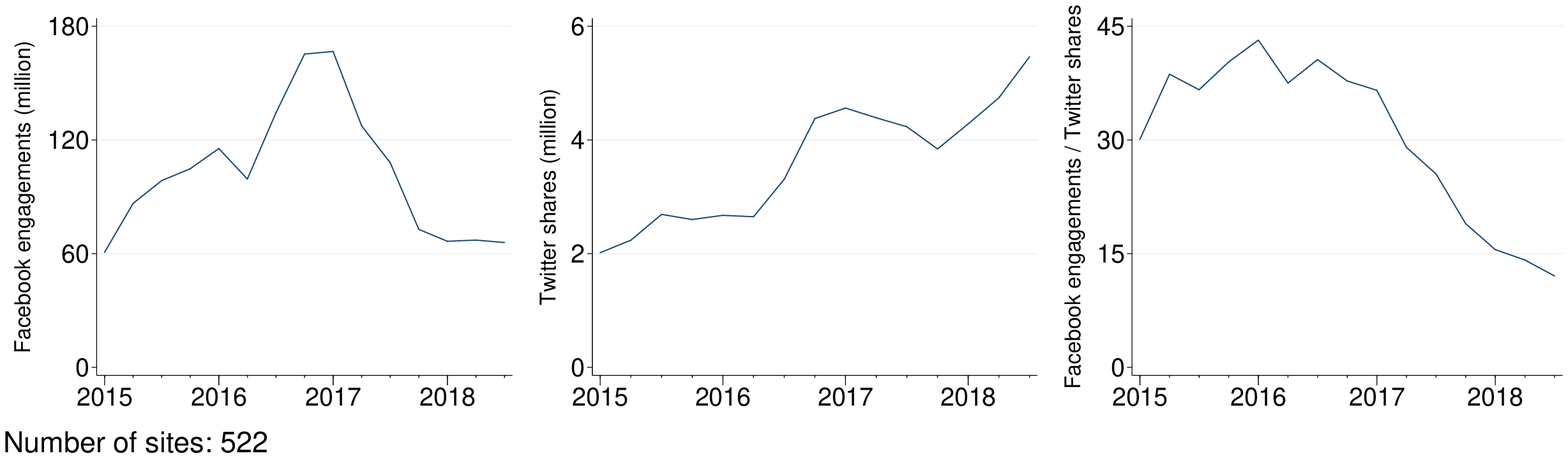}}\tabularnewline
\emph{Panel E: Excluding Sites Only Identified by FactCheck}\tabularnewline
{\footnotesize{}\includegraphics[scale=0.65]{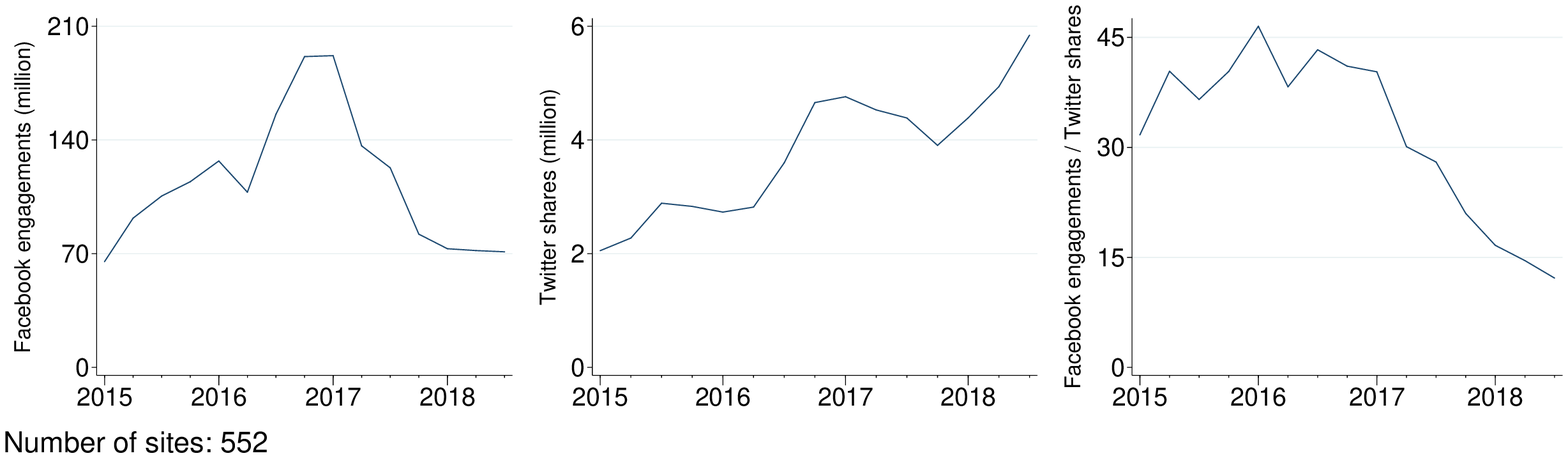}}\tabularnewline
\end{tabular}{\footnotesize\par}
\par\end{centering}
\medskip{}

\emph{\footnotesize{}Notes: }{\footnotesize{}This figure plots robustness
checks for the sample of fake news sites by excluding sites only identified
by a particular list. Each panel plots monthly Facebook engagements,
Twitter shares, and the ratio of Facebook engagements over Twitter
shares averaged by quarter. Panel A excludes sites only identified
by Grinberg er al. (2018). Panel B excludes sites only identified
by PolitiFact. Panel C excludes sites only identified by BuzzFeed.
Panel D excludes sites only identified by Guess et al. (2018). Panel
E excludes sites only identified by FactCheck.}{\footnotesize\par}
\end{figure}

\newpage{}

\begin{figure}[H]
{\footnotesize{}\caption{Robustness Checks of Fake News Sites - Time Coverage\label{fig:time-coverage}}
}{\footnotesize\par}

{\footnotesize{}\medskip{}
}{\footnotesize\par}
\begin{centering}
{\footnotesize{}}%
\begin{tabular}{c}
\emph{Panel A: Sites that Started Active Operation after November
2016}\tabularnewline
{\footnotesize{}\includegraphics[scale=0.65]{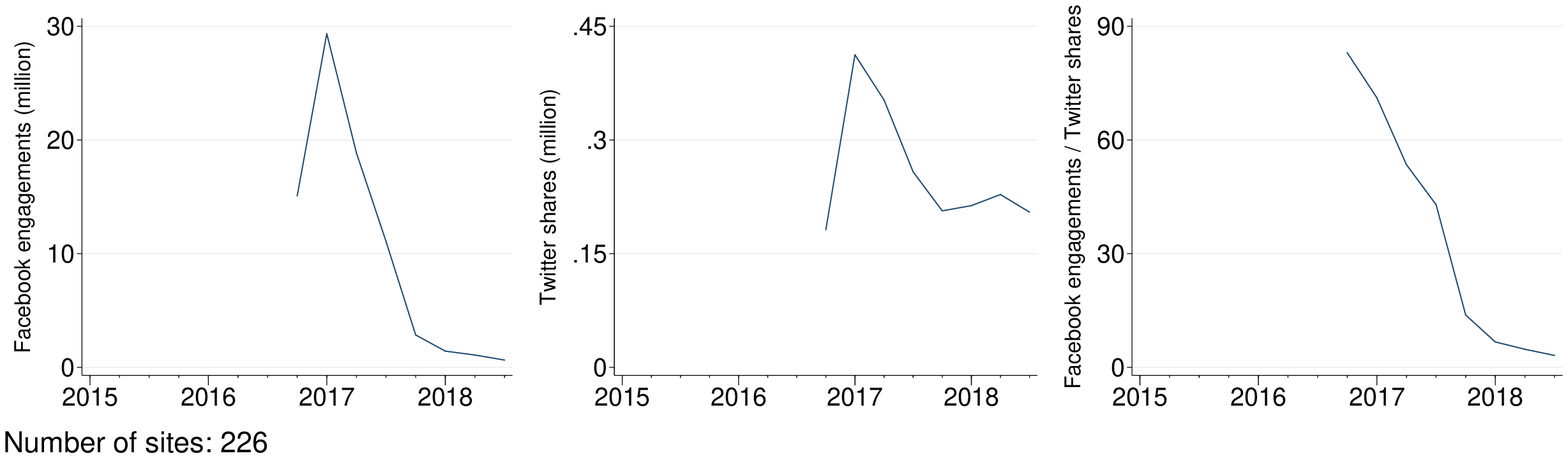}}\tabularnewline
\emph{Panel B: Sites that were in Active Operation in July 2018}\tabularnewline
{\footnotesize{}\includegraphics[scale=0.65]{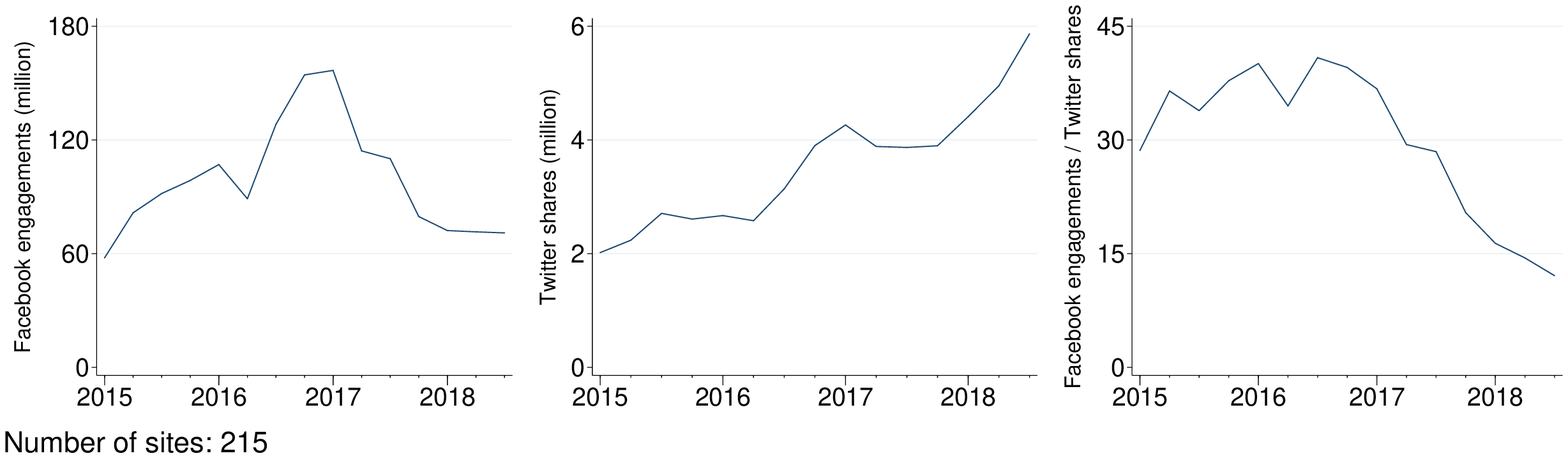}}\tabularnewline
\emph{Panel C: Sites that were in Active Operation during August 2015
to July 2018}\tabularnewline
{\footnotesize{}\includegraphics[scale=0.65]{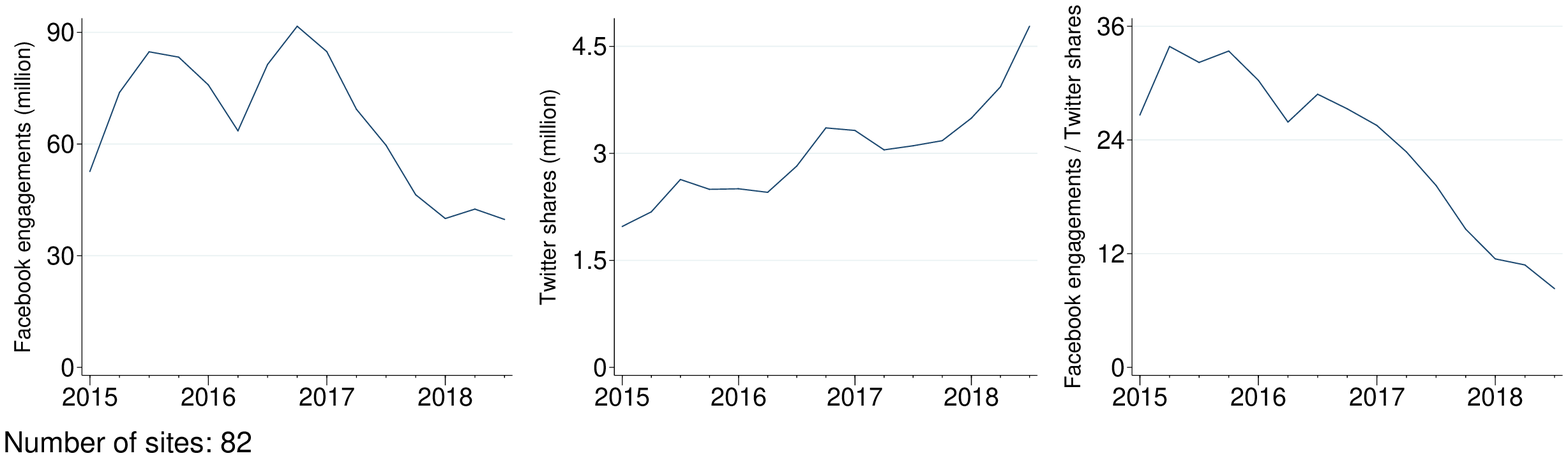}}\tabularnewline
\end{tabular}{\footnotesize\par}
\par\end{centering}
\medskip{}

\emph{\footnotesize{}Notes: }{\footnotesize{}This figure plots robustness
checks for the sample of fake news sites by looking at sites active
in different period. Each panel plots monthly Facebook engagements,
Twitter shares, and the ratio of Facebook engagements over Twitter
shares averaged by quarter. Panel A includes sites that started active
operation after the election in November 2016. Panel B includes sites
that were still in active operation in July 2018. Panel C includes
sites that were in active operation during August 2015 to July 2018.
A site is defined as being in active operation if it is tracked by
Alexa with a global rank higher than one million in terms of total
traffic.}{\footnotesize\par}
\end{figure}

\newpage{}

\begin{figure}[H]
{\footnotesize{}\caption{Robustness Checks of Fake News Sites - Number of Interactions\label{fig:number-of-interactions}}
}{\footnotesize\par}

{\footnotesize{}\medskip{}
}{\footnotesize\par}
\begin{centering}
{\footnotesize{}}%
\begin{tabular}{c}
\emph{Panel A: Excluding Top Five Fake News Sites}\tabularnewline
{\footnotesize{}\includegraphics[scale=0.65]{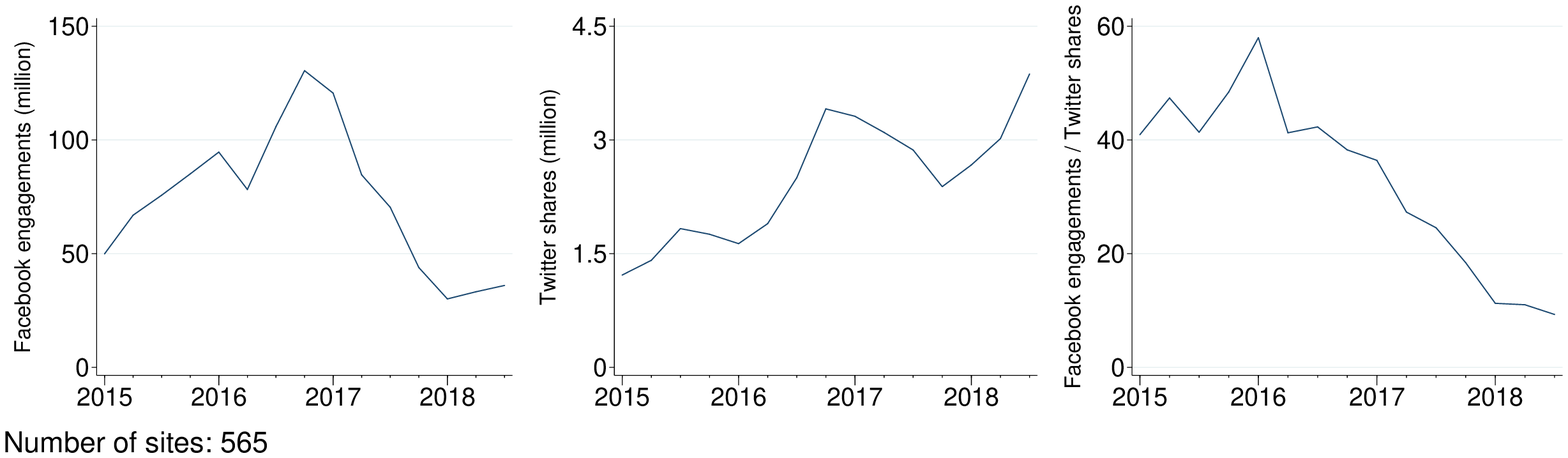}}\tabularnewline
\emph{Panel B: The First Decile of Fake News Sites}\tabularnewline
{\footnotesize{}\includegraphics[scale=0.65]{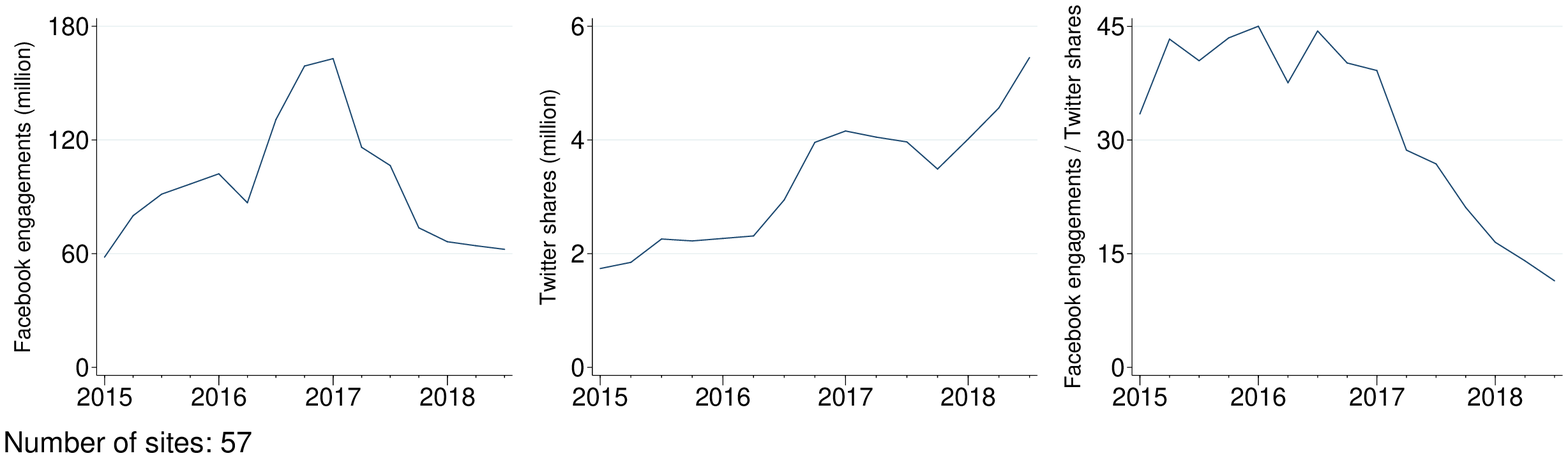}}\tabularnewline
\emph{Panel C: The Bottom Nine Deciles of Fake News Sites}\tabularnewline
{\footnotesize{}\includegraphics[scale=0.65]{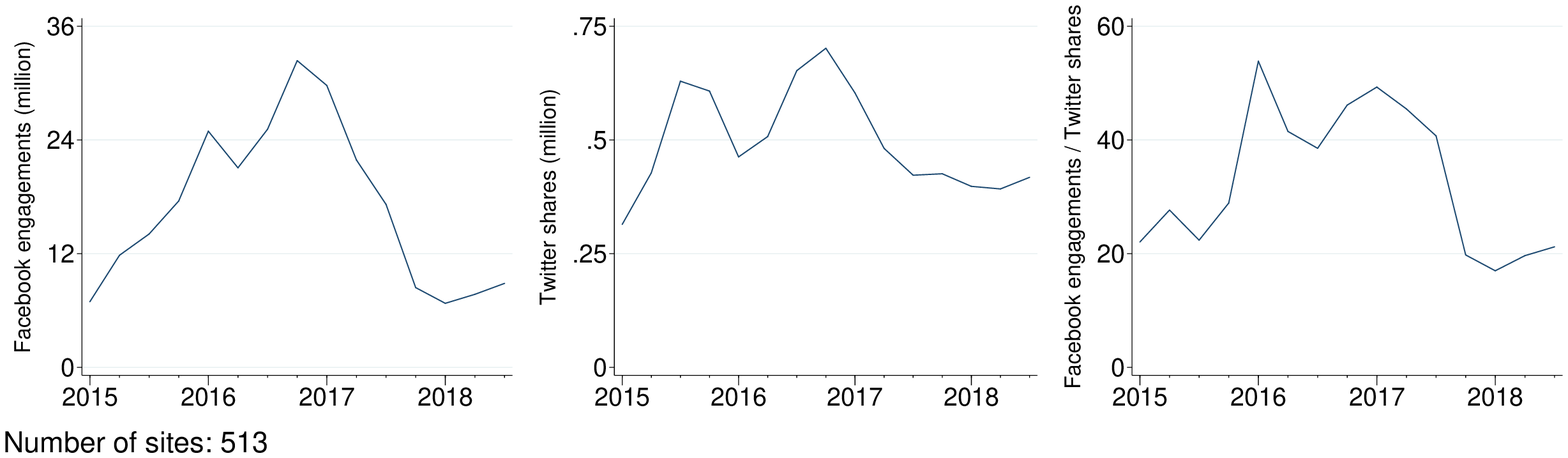}}\tabularnewline
\end{tabular}{\footnotesize\par}
\par\end{centering}
\medskip{}

\emph{\footnotesize{}Notes: }{\footnotesize{}This figure plots robustness
checks for the sample of fake news sites by excluding the largest
sites and looking at sites of different sizes. Each panel plots monthly
Facebook engagements, Twitter shares, and the ratio of Facebook engagements
over Twitter shares averaged by quarter. Panel A excludes five largest
sites in terms of total Facebook engagements plus Twitter shares from
January 2015 to July 2018. Panel B includes sites in the first decile.
Panel C includes sites in the bottom nine deciles, i.e., excludes
sites in the first decile. The deciles are also defined in terms of
total Facebook engagements plus Twitter shares during the sample period.}{\footnotesize\par}
\end{figure}

\newpage{}

\begin{figure}[H]
{\footnotesize{}\caption{Robustness Checks of Fake News Sites - Likelihood to Publish Misinformation\label{fig:likelihood}}
}{\footnotesize\par}

{\footnotesize{}\medskip{}
}{\footnotesize\par}
\begin{centering}
{\footnotesize{}}%
\begin{tabular}{c}
\emph{Panel A: Black Domains}\tabularnewline
{\footnotesize{}\includegraphics[scale=0.65]{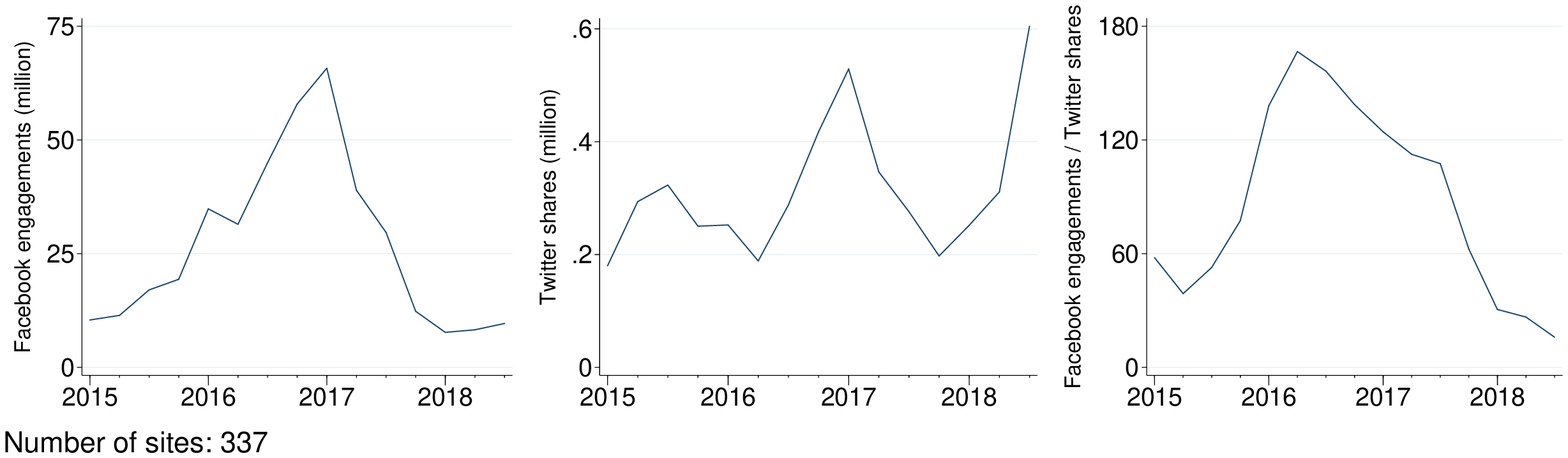}}\tabularnewline
\emph{Panel B: Red Domains}\tabularnewline
{\footnotesize{}\includegraphics[scale=0.65]{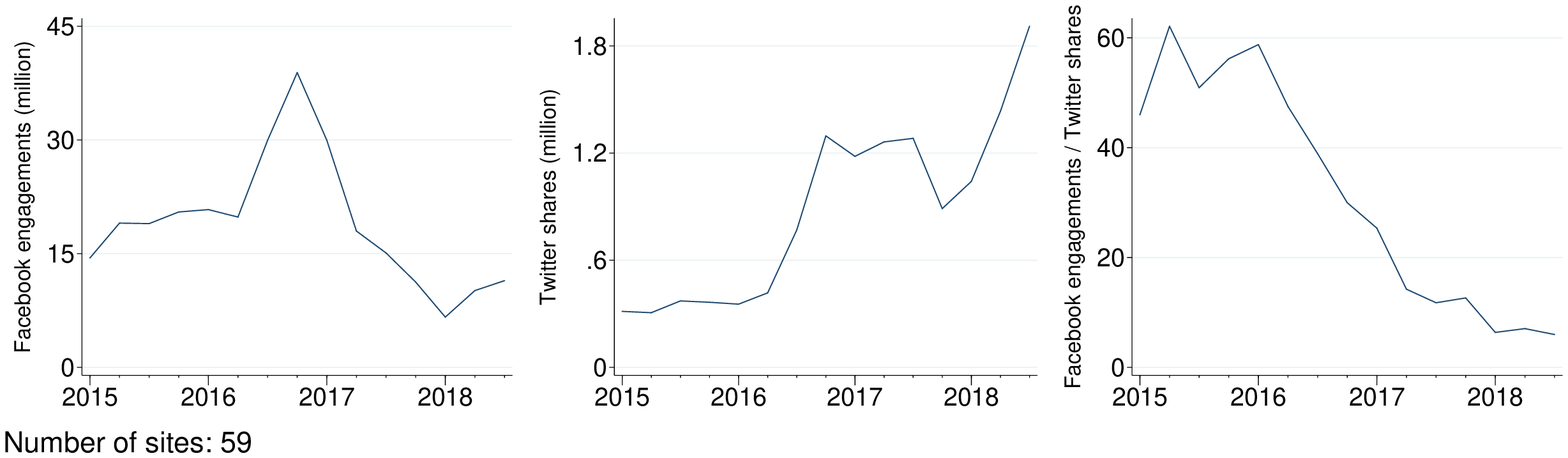}}\tabularnewline
\emph{Panel C: Orange Domains}\tabularnewline
{\footnotesize{}\includegraphics[scale=0.65]{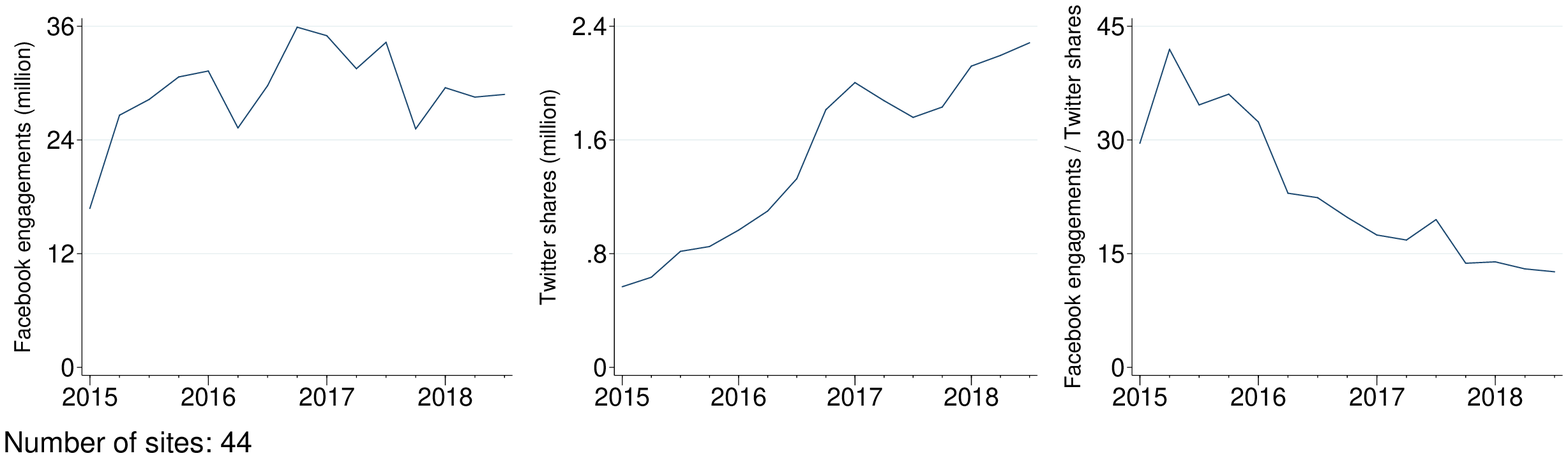}}\tabularnewline
\end{tabular}{\footnotesize\par}
\par\end{centering}
\medskip{}

\emph{\footnotesize{}Notes: }{\footnotesize{}This figure plots robustness
checks for three lists of fake news sites in Grinberg er al. (2018)
separately, classified by their likelihoods to publish misinformation.
Each panel plots monthly Facebook engagements, Twitter shares, and
the ratio of Facebook engagements over Twitter shares averaged by
quarter. The black domains were reported to have published entirely
fabricated stories, taken from pre-existing lists of fake news constructed
by the fact-checking and journalistic outlets Politifact, FactCheck,
and Buzzfeed, as well as domains used in other academic work. The
red and orange domains are identified by Snopes as sources of fake
news or questionable claims and classified by the authors by their
levels of perceived likelihood to publish misinformation: stories
from red domains have an extremely high likelihood of containing misinformation,
and stories from orange domains a high likelihood.}{\footnotesize\par}
\end{figure}
\newpage{}

\end{document}